\documentclass[fleqn,usenatbib]{mnras}

\usepackage{newtxtext,newtxmath}
\usepackage{pifont}
\usepackage[T1]{fontenc}

\DeclareRobustCommand{\VAN}[3]{#2}
\let\VANthebibliography\thebibliography
\def\thebibliography{\DeclareRobustCommand{\VAN}[3]{##3}\VANthebibliography}

\usepackage{sty/support-caption}
\usepackage{subcaption} 
\usepackage{graphicx}	
\usepackage{amsmath}	
\usepackage{xcolor}     

\usepackage{comment}
\def\ramses    {{\sc ramses}}
\def\sphinx    {{\sc sphinx}}
\def\galprop    {{\sc galprop}}
\def\ramsesrt    {{\sc ramses-rt}}
\def\numpy    {{\sc numpy}}
\def\matplotlib    {{\sc matplotlib}}
\def\pymses    {{\sc pymses}}
\def\adaptahop   {{\sc adaptahop}}
\def\Jlength {{\lambda_{\rm J, turb}}}
\def\DDtw {{\sc DD20}}

\newcommand\joki{\textcolor{red}}
\newcommand\marion{\textcolor{brown}}
\newcommand{\yo}[1]{{\color{blue}{{[YD: \bf #1]}}}}

\usepackage{soul}

\DeclareRobustCommand{\VAN}[3]{#2}

\title[CR feedback in disc galaxies]{Radiation-MagnetoHydrodynamics simulations of cosmic ray feedback in disc galaxies}

\author[Farcy et al.]{
Marion Farcy,$^{1}$\thanks{E-mail: marion.farcy@univ-lyon1.fr}
Joakim Rosdahl,$^{1}$
Yohan Dubois,$^{2}$
Jérémy Blaizot,$^{1}$
Sergio Martin-Alvarez$^{3}$
\\
$^{1}$Centre de Recherche Astrophysique de Lyon, CNRS UMR 5574, Univ. Lyon, Ens de Lyon, 9 avenue Charles André, F-69230 Saint-Genis-Laval, France\\
$^{2}$Institut d’Astrophysique de Paris, CNRS UMR 7095, UPMC Univ. Paris VI, 98 bis boulevard Arago, 75014 Paris, France\\
$^{3}$Institute of Astronomy and Kavli Institute for Cosmology, University of Cambridge, Madingley Road, Cambridge CB3 0HA, UK
}

\date{Accepted 2022 April 27. Received 2022 March 30; in original form 2022 February 2}

\pubyear{2022}

\begin{document}
\label{firstpage}
\pagerange{\pageref{firstpage}--\pageref{lastpage}}
\maketitle

\begin{abstract}
Cosmic rays (CRs) are thought to play an important role in galaxy evolution. We study their effect when coupled to other important sources of feedback, namely supernovae and stellar radiation, by including CR anisotropic diffusion and radiative losses but neglecting CR streaming. Using the \ramsesrt{} code, we perform the first radiation-magnetohydrodynamics simulations of isolated disc galaxies with and without CRs. We study galaxies embedded in dark matter haloes of $10^{10}$, $10^{11}$ and $10^{12}\, \rm M_{\odot}$ with a maximum resolution of $9 \,\rm pc$. We find that CRs reduce star formation rate in our two dwarf galaxies by a factor 2, with decreasing efficiency with increasing galaxy mass. They increase significantly the outflow mass loading factor in all our galaxies and make the outflows colder. We study the impact of the CR diffusion coefficient, exploring values from $\kappa = 10^{27}$ to $\rm 3\times 10^{29}\, cm^2\, s^{-1}$. With lower $\kappa$, CRs remain confined for longer on small scales and are consequently efficient in suppressing star formation, whereas a higher diffusion coefficient reduces the effect on star formation and increases the generation of cold outflows. Finally, we compare CR feedback to a calibrated 'strong' supernova feedback model known to sufficiently regulate star formation in high-redshift cosmological simulations. We find that CR feedback is not sufficiently strong to replace this strong supernova feedback. As they tend to smooth out the ISM and fill it with denser gas, CRs also lower the escape fraction of Lyman continuum photons from galaxies.
\end{abstract}



\section{Introduction}

The study of galaxy evolution is strongly related to the baryon cycle, which describes how gas collapses to form stars, and how stellar feedback then suppresses star formation and drives fountains of galactic outflows. Therefore, one of the key challenges of galaxy evolution is to understand the nature of the feedback processes that regulate star formation (SF) and gas expulsion, which in the end shape the galactic gas distribution at inter-stellar medium (ISM) and circum-galactic medium (CGM) scales.

It is commonly established that supernova (SN) feedback provides an important contribution in suppressing star formation and driving galactic winds, especially in low-mass galaxies \citep[e.g.][]{Dekel&Silk1986,Navarro&White1993,Gelli2020}. In the past, the ISM of galaxies could not be resolved in cosmological simulations, and it was beyond reach to model star formation and feedback from first principles. Instead, these processes were modelled with sub-grid recipes where SN feedback could be calibrated in various ways in order to reproduce a range of observations, such as the galaxy mass function and Kennicutt-Schmidt relation \citep[e.g.][]{Oppenheimer2010,Vogelsberger2014,Schaye2015}. In the last decade, however, it has become more and more feasible to resolve the ISM in simulations of galaxy evolution, opening the way for feedback (and star formation) models that are increasingly physically motivated and have less freedom for calibration \citep[see][for a recent review]{Vogelsberger2020}.

Several recent studies of galaxy evolution have applied this first principles approach to SN feedback \citep[e.g.][]{Smith2019,Peters2017,Hu2019, Fujimoto2019}. While SN feedback is found to have a strong impact on low-mass galaxies, they generally draw into question the assumption that this feedback process alone sufficiently suppresses star formation \citep[][and references hereafter]{Hopkins2014,Grudic2018,Smith2019}. Therefore, complementary feedback processes such as radiation feedback and cosmic rays are likely important, if sub-dominant.

Stellar radiation interacts with ISM gas, through photoionization heating of gas and radiation pressure \citep[e.g.][]{Hopkins2014, Peters2017, Emerick2018}. However, self-consistent radiation hydrodynamics simulation studies such as those from \citet{Rosdahl2015} and \citet{Kannan2019} find that photoionization heating has a non-negligible but insufficient effect in regulating star formation, and that radiation pressure only has a marginal effect.

Cosmic rays (CRs) have been proposed by many as an additional important source of feedback. When supernovae explode, the shock waves generated accelerate charged particles up to relativistic velocities through diffusive shock acceleration \citep{Axford1977,Krymskii1977,Bell1978,Blandford&Ostriker1978}. By nature, CRs have a number of advantages for being an efficient feedback source. Being at equipartition with magnetic, turbulent and gravitational energies \citep[][from measurements of the Milky Way]{Boulares&Cox1990}, they provide a significant non-thermal pressure that can drive the gas dynamics, on scales ranging from their injection sites to the CGM. They have a softer equation of state than the thermal energy, so their pressure drops less quickly upon adiabatic expansion. They cool less efficiently than non-relativistic gas \citep{Ensslin2007}, so their energy is maintained longer than the thermal energy of the gas. Additionally, a part of the CR energy lost through collisions and Coulomb interactions is delivered to the gas which is heated. These properties of CRs have been shown to suppress star formation \citep[e.g.][in idealised galaxies]{Jubelgas2008,Pfrommer2016,Chan2019,Semenov2021} and drive dense and cold winds in a number of studies (e.g. \citealp{Booth2013,Salem&Bryan2014,Pakmor2016,Wiener2017,Jacob2018,Dashyan&Dubois2020,Jana2020,Girichidis2021} in idealised galaxies, \citealp{Farber2018,Girichidis2018} in stratified boxes of ISM and \citealp{Buck2020,Hopkins2020,Ji2020,Butsky2021} in cosmological zoom-in simulations).

Therefore, CRs appear a promising complementary feedback mechanism to limit the growth of galaxies in the Universe. However, radiation, SN, and CR feedback have never been considered before in combination. Recently, \citet[\DDtw{} hereafter]{Dashyan&Dubois2020} studied CR feedback in two isolated disk galaxies spanning an order of magnitude in mass. However they did not consider radiation feedback and they used a fairly simple and locally inefficient model for star formation which does not represent the state-of-the-art used in recent cosmological simulations. We therefore expand on the work of \DDtw{} with the first Radiation-MagnetoHydroDynamics (RMHD) simulations of galaxy evolution combining ideal magnetohydrodynamics (MHD), SN feedback, radiative transfer and CRs to study the combined effect of these processes. Using the \ramsesrt{} code \citep{Teyssier2002,Teyssier2006,Rosdahl2013}, we investigate how CR feedback shapes galaxy growth, studying the effects of CRs in regulating star formation and the ISM and CGM gas of three idealised galaxies spanning two orders of magnitude in mass and with resolution down to 9 pc.

CR transport is a complex process that includes advection with gas, anisotropic diffusion and streaming down the CR pressure gradient. Depending on whether the sources of CR scattering are external magnetic field inhomogeneities or waves excited by CRs themselves, the importance of each of these processes can vary significantly and impact the diffusion coefficient, through which CR propagation is parameterized \citep[see e.g.][]{Zweibel2017}. While this parameter is poorly constrained, we know from other studies \citep[e.g.][]{Salem&Bryan2014,Farber2018,Jacob2018,Chan2019, Dashyan&Dubois2020,Hopkins2020, Jana2020, Girichidis2021, Semenov2021} that conclusions on the role of CRs as a feedback source can differ quite dramatically depending on its value. We therefore test the variability of CR feedback using five values of diffusion coefficient, pursuing the study initiated by \DDtw{}, with an increased sample of galaxies and with our physically motivated setup.

Our eventual goal is to determine if cosmic rays, combined with SN and stellar radiation, constitute a feedback model sufficient to regulate the growth of low-mass galaxies in the Universe. To circumvent the limited predictive power of our non-cosmological galaxy disk simulations, a preliminary way of answering this question is to compare this combined feedback with the artificially boosted SN feedback model previously used in the \sphinx{} cosmological simulations \citep{Rosdahl2018}, which is shown to sufficiently regulate star formation at high redshift to reproduce the observed galaxy luminosity function.

The structure of this paper is as follows. Section~\ref{section:simu} introduces the code, methods and setup used to perform our isolated disc simulations. Section~\ref{section:results} first focuses on the qualitative effects of CRs on our galaxies. In Section~\ref{subsection:sfr}, we investigate the efficiency of cosmic ray feedback in regulating star formation, before studying its effects on the mass loading factor and the temperature phases of the outflowing gas in Section~\ref{subsection:cgm}. In Section~\ref{subsection:G10_dcr}, we further explore the variability of our results when changing the cosmic ray diffusion coefficient, a key parameter governing their propagation and their role as a feedback source. We analyse to what extent CRs can shape galaxy evolution compared to a calibrated stronger SN feedback in Section~\ref{subsection:boostvscr}. We consider the consequences of those two feedback models on the escape of Lyman Continuum radiation in Section ~\ref{subsection:fesc}. We finally give an overview of the main results of this paper in the context of other studies in Section~\ref{section:discussion} and conclude in Section~\ref{section:ccl}.  

\section{Simulations and methods}
\label{section:simu}

To perform Radiation-MagnetoHydroDynamics simulations of isolated galaxies, we use the \ramsesrt{} adaptive mesh refinement (AMR) code \citep{Rosdahl2013,Rosdahl&Teyssier2015}, a radiation-hydrodynamics (RHD) extension of the \ramses{} code \citep{Teyssier2002}. The solver described by \citet{Fromang2006} is employed to compute the full set of ideal MHD equations. The fluxes are solved with the Harten-Lax-van Leer Discontinuities (HLLD) Riemann solver \citep{Miyoshi&Kusano2005} and the minmod total variation diminishing slope limiter \citep{vanLeer1979}. The magnetic field evolves following the induction equation, which is implemented using a constrained transport method, which ensures a null magnetic divergence by construction, and employs the second order Godunov scheme MUSCL \citep{Teyssier2006}. The radiative transfer equations are solved with a two-moment method and the M1 closure for the Eddington tensor. The code tracks the non-equilibrium ionization states of hydrogen and helium in each gas cell, and includes the effects of radiation pressure, photoheating and radiative cooling. Finally, we combine \ramsesrt{} with the method developed by \citet{Dubois&Commercon2016} to solve the anisotropic diffusion of CRs. We further add the minmod slope limiter on the transverse component of the flux that preserves the monotonicity of the solution in the asymmetric method of~\cite{Sharma&Hammett2007}, as described in \DDtw{}.

We simulate galaxy discs of baryonic mass $3.5\times 10^8$, $3.5\times 10^9$ and $3.5\times 10^{10} \rm \ M_{\odot}$ embedded in $10^{10}$, $10^{11}$ and $ 10^{12} \rm \ M_{\odot}$ dark matter haloes respectively. We refer to them as G8, G9 and G10, where the numbers stand for the order of magnitude of the galaxy baryonic mass.

\subsection{Galaxy disc setup}
\label{subsection:discs}

\begin{table*}
	\centering
	\caption{Main parameters of the three disc galaxies. From left to right: galaxy name (number connected to the disc mass), $M_{\rm disc}$: baryonic disc mass (gas + stars), $M_{\rm halo}$: dark matter halo mass, $R_{\rm vir}$: halo virial radius, $L_{\rm box}$: length of the simulated box, $\Delta x_{\rm max}$: maximum cell size, $\Delta x_{\rm min}$: minimum cell size, $m_{*}$: stellar particle mass, $f_{\rm gas}$: gas disc fraction, $Z_{\rm disc}$: disc metallicity, $t_{\rm end}$: time reached at the end of the run, for the last snapshot.}
	\label{tab:run_prop}
	\begin{tabular}{cccccccccccc}
		\hline
		Galaxy & \vline & $M_{\rm disc}$ &  $M_{\rm halo}$ & $R_{\rm vir}$ & $L_{\rm box}$ & $\Delta x_{\rm max}$ & $\Delta x_{\rm min}$ & $m_{*}$ & $f_{\rm gas}$ & $Z_{\rm disc}$ & $t_{\rm end}$          \\
		name   & \vline & [$\rm M_{\odot}$] & [$\rm M_{\odot}$] & [kpc] & [kpc] & [kpc] & [pc] & [$\rm M_{\odot}$] & & [$\rm Z_{\odot}$] & [Myr] \\
		\hline
		G8     & \vline & $3.5\times10^{8}$ & $10^{10}$ & 41 & 150 & 2.34 & 9 & 2500 & 0.5 & 0.1 & 500 \\
		G9     & \vline & $3.5\times10^{9}$ & $10^{11}$ & 89 & 300 & 2.34 & 9 & 2500 & 0.5 & 0.1 & 500 \\
		G10    & \vline & $3.5\times10^{10}$ & $10^{12}$ & 192 & 600 & 4.68 & 18 & 20000 & 0.3 & 1 & 500 \\
		\hline
	\end{tabular}
\end{table*}

The initial conditions for all our simulations are generated using the \textsc{Makedisc} code \citep{Springel2005}. A more complete description can be found in \citet{Rosdahl2015}, and some of the main properties of the discs are summarised in Table~\ref{tab:run_prop}. Each of our three discs is hosted in a dark matter (DM) halo which follows a NFW density profile \citep{NFW1997}, with a concentration parameter $c = 10$ and a spin parameter $\lambda = 0.04$. The DM is modelled by collisionless particles all of the same mass, $10^5$  particles for G8 and $10^6$ of them for G9 and G10, leading to a DM particles mass of $\rm 10^{5}\ M_{\odot}$ for G8 and G9 and $\rm 10^{6}\ M_{\odot}$ for G10. The discs also have an initial distribution of gas and stellar particles, both following an exponential density profile in radius (with a scale radius of 0.7, 1.5 and 3.2 kpc by increasing order of galaxy mass) and a Gaussian in height (with the scale height being one tenth of the scale radius). Initially, the disc gas has a uniform temperature of $T=10^4$ K while the rest of the box is filled with a diffuse circum-galactic gas at $10^6$ K, and a hydrogen density $ n_{\rm H}=10^{-6}\ \rm cm^{-3}$. The metallicity of the gas disc is set to 0.1 $Z_{\odot}$\footnote{We assume in this work a Solar metal mass fraction of $Z_{\odot}=0.02$} for both G8 and G9 and to 1 $Z_{\odot}$ for G10, and the CGM metallicity is set to zero. This setup describes an idealised CGM, initially almost empty from gas, and which is not designed to be realistic. We note that this description of the CGM is very simplified and does not represent very well, especially not initially, the CGM found in cosmological simulations, populated with a multi-phase mix of inflowing and outflowing gas. The initial stellar particles do not explode as SN, nor provide any other feedback to the surrounding gas. They account for 50\% of the total initial baryonic mass of the discs for G8 and G9 galaxies and 70\% for G10. 10\% of the stellar particles are distributed in a stellar bulge and the remainder throughout the disc according to the gas profile described above, so that the bulge to total (disc plus bulge) stellar mass ratio is 0.1.

\subsection{Adaptive refinement}
\label{subsection:AMR}

The \ramses{} code uses an adaptive refinement scheme, where each cell can be divided into 8 children cells of width half that of the parent. Equivalently, this means that the size of a cell refined at a level $\ell$, $\Delta x_{\ell}$, is twice smaller than the size of the next coarser cell of level $\ell-1$, so that $\Delta x_{\ell} = {L_{\rm box}}$ $/$ $2^{\ell}$, with $L_{\rm box}$ being the full size of the simulation box. We flag a cell to be refined if its total mass (dark matter and baryons) is higher than the mass of 8 dark matter particles (which corresponds to $8\times 10^5\ \rm M_\odot$ for G8 and G9 and $8\times 10^6\ \rm M_\odot$ for G10), or if its width is larger than a quarter of the local Jeans length. In this study, the three disc galaxies G8, G9 and G10 are located at the centres of boxes of 150, 300 and 600 kpc in width respectively. We adopt a maximum cell resolution of $\Delta x_{\rm max}=\rm 9\ \rm{pc}$ for G8 and G9, but 18 pc for G10. The minimum cell resolution is $\Delta x_{\rm min}=\rm 2.34\ {\rm kpc}$ for G8 and G9, and 4.68 kpc for G10. We briefly discuss resolution convergence in Section \ref{section:discussion}.

\subsection{Radiative transfer}
\label{subsubsection:rt}

The radiative transfer equations in \ramsesrt{} are solved with a first-order moment method, using the M1 closure relation for the Eddington tensor, and the Global Lax–Friedrichs (GLF) intercell flux function for the advection of the photon fluids (see \citealp{Rosdahl2013}). To reduce the computational cost of light propagation, we use a reduced speed of light of $c/100$.
We solve the non-equilibrium chemistry and radiative cooling of neutral and ionized hydrogen and helium, for which we follow the ionization fractions. For the three photon groups (HI, HeI and HeII ionizing photons), we adopt a dust absorption opacity of $\rm 10^3\ cm^2\, g^{-1}$ ($Z/\rm Z_{\odot}$). As listed in Table~\ref{tab:photon_list}, each photon group is defined by a frequency interval, for which we track photon density and flux in each cell. Stars emit photons at a rate derived from version 2.2.1 of the Binary Population And Spectral Synthesis model \citep[BPASS;][]{Stanway2016, Stanway&Eldridge2018}. We assume an initial mass function close to \citet{Kroupa2001} with slopes of -1.3 from 0.1 to 0.5 $\rm M_{\odot}$ and -2.35 from 0.5 to 100 $\rm M_{\odot}$. Atomic metal cooling for gas with temperature $T>10^4$ K is computed using cooling rates tabulated from \textsc{cloudy} \citep{Ferland1998}, and fine-structure line cooling is enabled for gas with $T < 10^4$ K, using the fitting function from \citet{Rosen&Bregman1995}.
We also include gas heating from an external redshift zero uniform UV background, following \citet{Haardt&Madau2012}, with self-shielding for $n_{\rm H}>10^{-2}\ \rm H\ cm^{-3}$.

\begin{table}
	\centering
	\caption{Properties of the three photon groups used in this study. From left to right: photon group name, $\epsilon_0$ and $\epsilon_1$: minimum and maximum photon energy range, $\overline{\epsilon}$: mean photon energy $\pm 10\%$.}
	\label{tab:photon_list}
	\begin{tabular}{ccccc}
		\hline
		Photon group & \vline & $\epsilon_0$ [eV] & $\epsilon_1$ [eV] & $\overline{\epsilon}$ [eV]\\
		\hline
		UV$\rm _{HI}$ & \vline & 13.60 & 24.59 & 18 \\
		UV$\rm _{HeI}$ & \vline & 24.59 & 54.42 & 33.4 \\
		UV$\rm _{HeII}$  & \vline & 54.42 & $\infty$ & 60 \\
		\hline
	\end{tabular}
\end{table}

\subsection{Star formation}
\label{subsection:SF}

We turn gas into star particles only if cells at the highest level of refinement are gravitationally unstable, i.e. if they have a width larger than the turbulence Jeans length defined as: 
\begin{equation}
    \Jlength = \frac{\pi\sigma_{\rm gas}^2 + \sqrt{36\pi c_s^2G\Delta x^2\rho + \pi^2\sigma_{\rm gas}^4}}{6G\rho\Delta x}\, ,
\end{equation}
where $G$ is the gravitational constant, $\sigma_{\rm gas}$ is the gas velocity dispersion computed using the velocity gradients with neighbour cells, $c_s$ is the local sound speed and $\rho$ is the gas density. We note that neither the magnetic nor the cosmic ray pressure contribute to the sound speed in the calculation of the Jeans length.

Gas is converted into stars at a rate:
\begin{equation}
    \dot{\rho}_* = \epsilon \rho / t_{\rm ff}\, ,
    \label{eq:SF_conv}
\end{equation}
where $\epsilon$ is the star formation efficiency and $t_{\rm ff} = (3\pi/(32 G\rho))^{1/2}$ is the gas free-fall time. Stellar populations are represented by collisionless stellar particles with an initial mass which is an integer multiple of $m_*$, whose value varies with galaxy mass and is listed in Table~\ref{tab:run_prop}. The conversion from gas to stars is done by stochastically sampling a Poisson mass-probability distribution, as detailed by \citet{Rasera&Teyssier2006}, so that the conversion rate described in Eq.~\ref{eq:SF_conv} holds only on average. 

We do not use a global constant star formation efficiency but rather a local $\epsilon$ depending on the gravo-turbulent properties of the gas, based on the work of \citet{Federrath&Klessen2012} (for details, see \citealp{Kimm2017} or \citealp{Trebitsch2017}).
We show in Appendix~\ref{app:dens-fk2} that the highly varying local star formation efficiency tends to create a bursty and clumpy star formation compared to the more widely used constant and small $\epsilon$. Consequently, we expect stronger and more localised feedback events compared to what is found by \DDtw{} who form stars with a constant 2\% efficiency if the hydrogen density in the cell is $n_{\rm H} \geq 10^2 \rm \, H\, cm^{-3}$.



\subsection{Stellar feedback}
\label{subsection:stellar-feedback}
We include stellar feedback in the form of type II supernova explosions, photoionization, photoheating and radiation pressure. We use the mechanical feedback prescription of \citet{Kimm&Cen2014} and \citet{Kimm2015} to deposit momentum in the cells neighbouring SN explosions. Considering the local simulation resolution and the gas density and metallicity, this method adapts the radial momentum depending on how well the Sedov-Taylor phase is resolved. Doing so, we limit the numerical radiative losses due to a lack of resolution.

Following this prescription, each stellar particle explodes in multiple events between 3 and 50 Myr after its birth, each explosion releasing an energy $E_{\rm SN}=10^{51}$ ergs. This is another difference between our setup and that of \DDtw{}, in which a stellar particle explodes in one single cumulative event 5 Myr after its formation. The number of explosions $N_{\rm  SN}$ per particle is defined as:
\begin{equation}
    N_{\rm SN}=\frac{m_*\eta_{\rm SN}}{M_{\rm SN}} \,
\end{equation}
where $m_*$ is the stellar particle mass, $\eta_{\rm SN}$ is the mass fraction of the stellar population exploding as type II SNe, and $M_{\rm SN}$ is the average mass of those exploding stars. We assume a Kroupa Initial Mass Function (IMF), following which we adopt $\eta_{\rm SN}=0.2$ and $M_{\rm SN}=19.1\ \rm M_\odot$.


For our runs including CR feedback, we take 10\% of the energy otherwise released with each SN explosion and instead release it into the host cell in the form of CR energy\footnote{The cosmic ray energy injection does not contribute to the thermal momentum injection because, unlike the thermal pressure, the CR pressure does not substantially cool down over one time step at any of the gas densities sampled in our simulations, and the build up of momentum by CR pressure is always resolved \citep[but see also][]{Diesing&Caprioli2018,Curro2021}.}. The 10\% value is commonly used in simulations of CR feedback and is suggested by observations of local supernova remnants \citep{Hillas2005, Strong2010, Morlino&Caprioli2012,Dermer&Powale2013}. We provide more details on the equations at stake in those energy exchanges in Section~\ref{subsection:MHD}.



\subsection{Magnetic field and cosmic ray propagation}
\label{subsection:MHD}

Following \DDtw{}, we initialise our simulations with a toroidal magnetic field permeating the disc of our galaxies, reproducing the large-scale field observed in galaxies \citep{Beck2015}. To ensure that the divergence of the magnetic field $\bmath{B}$ cancels we initialise this toroidal magnetic field as the curl of a vector potential $\bmath{A}$ set to:
\begin{equation}
    \bmath{A} = \frac{3}{2}B_0r_0\left(\frac{\rho}{\rho_0}\right)^{\frac{2}{3}}\ \bmath{e_z} \, ,
	\label{eq:A}
\end{equation}
where $\rho$ corresponds to the gas density profile, $\rho_0$ its normalisation of $\sim 15 \ \text{cm}^{-3}$ (for G8, G9 and G10) and $r_0$ its scale radius of 3.2 kpc for G10, 1.5 kpc for G9 and 0.7 kpc for G8. $\bmath{e_z}$ is the z-axis unit vector in a Cartesian coordinate system. The initial magnetic field strength $B_0$ is set to $\rm 1 \mu G$.

CRs are advected by the bulk motion of the gas and diffused along the magnetic field, following the advection-diffusion approximation described by \citet{Dubois&Commercon2016} and \citet{Dubois2019}. Physically, CRs are highly energetic charged particles whose motion is thus strongly restricted to the surrounding magnetic field. In \ramses{}, we consider CRs as a relativistic fluid with an adiabatic index $\rm \gamma_{CR}=4/3$ and tracked through a non-thermal pressure term. CRs diffuse along magnetic field lines with a fiducial diffusion coefficient $\rm \kappa=10^{28}\ cm^2\,s^{-1}$, as determined to correspond to collisionless particles of a few GeV where most of CR energy density resides \citep{Strong2007, Trotta2011}. Including the CR contribution, the total energy of the fluid is:

\begin{equation}
    e = \frac{\rho u^2}{2}+e_{\rm th}+e_{\rm CR}+\frac{B^2}{8\pi}\, ,
\end{equation}
where $e_{\rm th}$ and $e_{\rm CR}$ are respectively the thermal and CR energy per unit of volume contained in one cell, and $u$ is the gas velocity. The evolution of the different energy and the magnetic field are described by the following MHD equations, in the framework of ideal MHD:

\begin{align}
    \label{eq:induction}
    &\frac{\partial\bmath{B}}{\partial t} = \bmath{\nabla} \times
    \left( \bmath{u} \times \bmath{B} \right)\\
    \label{eq:cons}
    &\frac{\partial\rho}{\partial t} + \bmath{\nabla} \cdot
    \left( \rho \bmath{u} \right) = 0\\
    \label{eq:cons-mom}
    &\frac{\partial\rho\bmath{u}}{\partial t} + \bmath{\nabla} \cdot
    \left( \rho \bmath{u}\bmath{u} + P_{\rm tot} - \frac{\bmath{B}\bmath{B}}{4\pi}\right) = \rho\bmath{g}\\
    \label{eq:etot}
    &\frac{\partial e}{\partial t} + \bmath{\nabla} \cdot
    \left( (e+P_{\rm tot})\bmath{u} - \frac{\bmath{B}(\bmath{B}\cdot \bmath{u})}{4\pi} \right) = \rho\bmath{u}\cdot\bmath{g} + Q_{\rm CR} + Q_{\rm th}\nonumber\\
    &- \Lambda_{\rm rad} - \Lambda_{\rm CR} - \bmath{\nabla}\cdot\bmath{F_{\rm CR}}\\
    \label{eq:ecr}
    &\frac{\partial e_{\rm CR}}{\partial t} + \bmath{\nabla} \cdot
    \left(e_{\rm CR}\bmath{u} \right) = -P_{\rm CR}\bmath{\nabla}\cdot\bmath{u} + Q_{\rm CR} - \Lambda_{\rm CR} - \bmath{\nabla}\cdot\bmath{F_{\rm CR}}
\end{align}

In these equations, the total pressure $P_{\rm tot}=P_{\rm th} + P_{\rm CR} + P_{\rm mag}$ where the magnetic pressure $P_{\rm mag}=B^2/(8\pi)$, the CR pressure $P_{\rm CR}=e_{\rm CR}(\gamma_{\rm CR}-1)$, and the thermal pressure $P_{\rm th}=e_{\rm th}(\gamma-1)$, with $\gamma_{\rm CR}$ and $\gamma$ the adiabatic indices for CRs and gas. We assume a purely monoatomic gas with $\gamma=5/3$. Among the other quantities, $\bmath{g}$ is the gravitational field, and $Q_{\rm th}$ and $Q_{\rm CR}$ are respectively thermal and CR energy source terms and contribute to the gas heating, with the former including heating from the UV background and CR collisional heating. $\Lambda_{\rm rad}$ and $\Lambda_{\rm CR}$ are cooling terms representing radiative and CR energy losses respectively. We note that the $\Lambda_{\rm CR}$ component is due to Coulomb and hadronic collisions from which a reinjection to the thermal component is already taken into account in the $Q_{\rm th}$ term \citep{Guo&Oh2008}. The anisotropic diffusion flux term is $\bmath{F_{\rm CR}}=-\kappa\bmath{b}\left(\bmath{b}\cdot\bmath{\nabla}e_{\rm CR}\right)$ with $\bmath{b}=\bmath{B}/||\bmath{B}||$ the magnetic field unit vector.
The streaming terms, which introduce a transfer of energy from CR pressure to thermal pressure, and an advection term at about the Alfv\'en velocity, are neglected in this work, as they have high computational cost and were found by \DDtw{}, with a very similar setup, to have secondary effects on the gas dynamics. Since we have similar resolution and ISM structure in our simulations as \DDtw{}, we disregard CR streaming.

\section{Results}
\label{section:results}

Throughout this section, we denote simulations with and without CRs as 'CR' and 'noCR'. We first provide a qualitative comparison of the discs with and without CR feedback.


\begin{figure}
    \centering
	\includegraphics[width=\columnwidth]{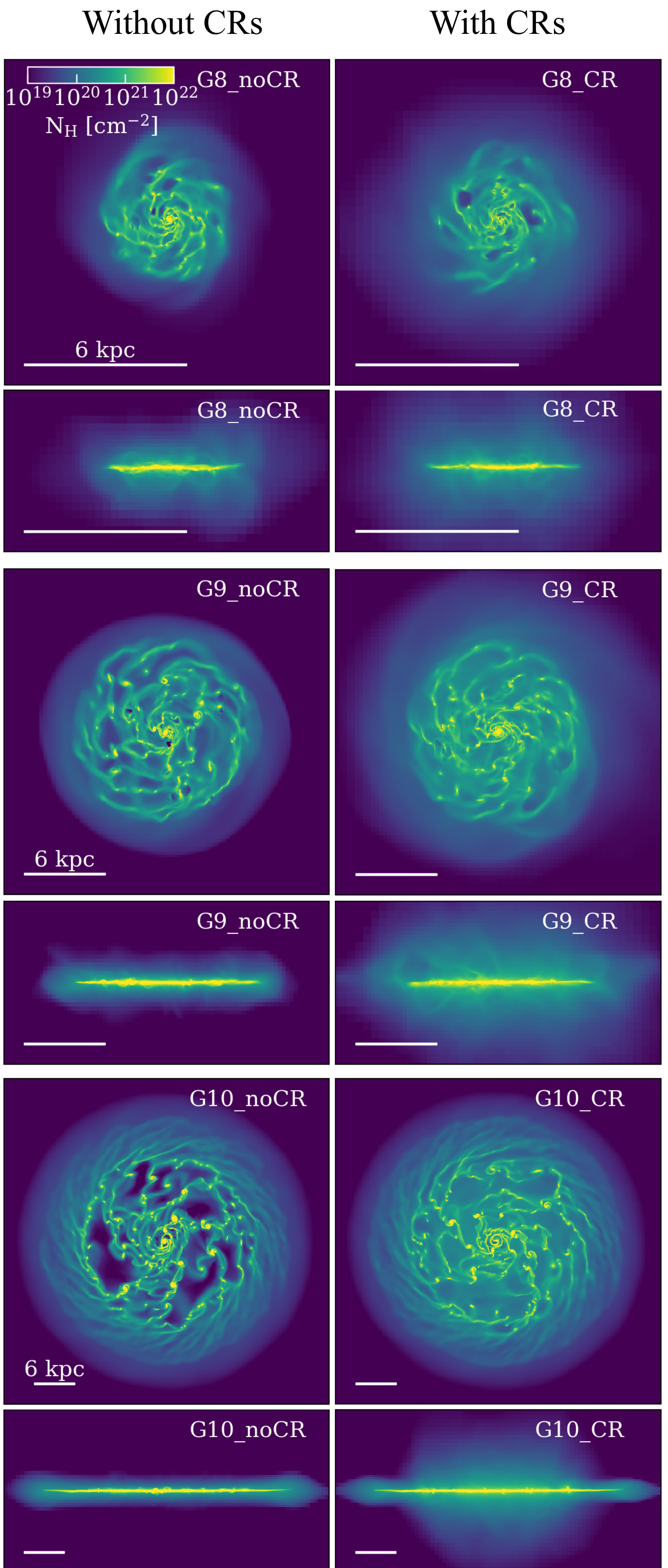}
    \caption{Maps of the three discs at 500 Myr in order of increasing mass from top to bottom. Respectively for each galaxy, 12, 24 and 48 kpc maps of face-on and edge-on hydrogen column density are shown, for the noCR discs in the left column and with CRs added on the right. The name of each run is written in the upper right corner of the maps, and a 6 kpc width scale bar is plotted in the lower left corner of each panel. The three discs tend to be thicker and with a smoother gas distribution when CRs are included.}
    \label{fig:nHmaps}
\end{figure}

\begin{figure}
    \centering
	\includegraphics[width=\columnwidth]{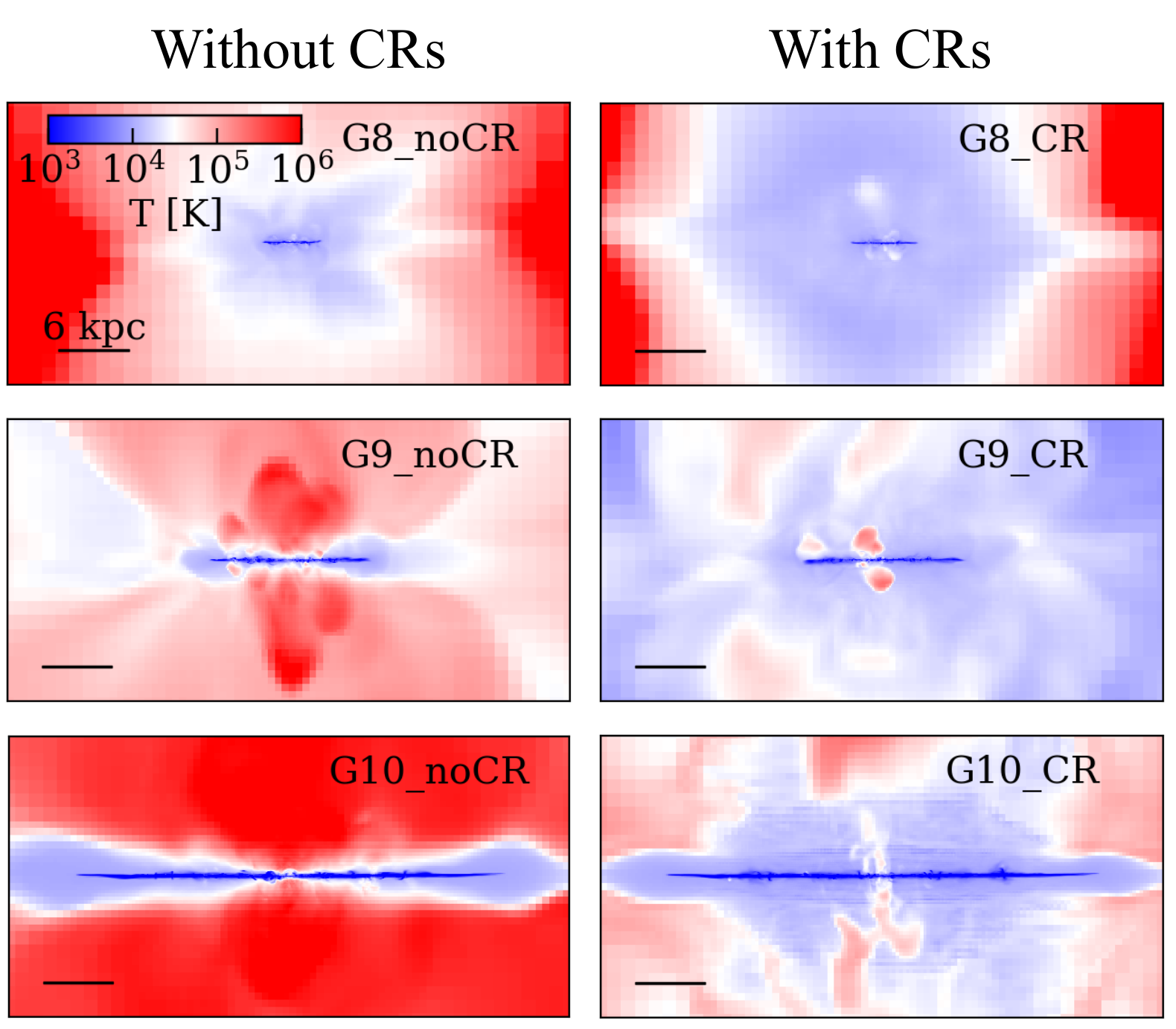}
    \caption{Mass-weighted 48 kpc-wide slices of the three discs at 500 Myr in order of increasing mass from top to bottom. For each galaxy, edge-on temperature maps are plotted for the noCR case in the left column and with CRs added on the right. The circum-galactic medium of the galaxies becomes much colder with CRs included.}
    \label{fig:Tmaps}
\end{figure}

Fig.~\ref{fig:nHmaps} shows face-on and edge-on maps of the hydrogen column density for the three discs, comparing runs without (left) and with (right) CR feedback. In the face-on maps, one can see clumps of dense gas, which are sites of star formation. 

Comparing the left and right panels, we see that the CR feedback tends to smooth out the ISM\footnote{For this qualitative analysis, we somewhat arbitrary define the ISM as being gas within 1 kpc from the disc plane.} in all our simulated galaxies, producing a more extended and diffuse gas distribution. By the end of our runs, the gas disc is thicker at any galaxy mass when CRs are included, as seen in the edge-on hydrogen density maps, in agreement with e.g. \citet{Salem2016} and \citet{Buck2020}. 

The CR feedback produces not only denser but also colder gas in the vicinity of the ISM. This is visible in Fig.~\ref{fig:Tmaps} showing edge-on temperature maps\footnote{Note that the temperature maps in Fig.~\ref{fig:Tmaps} are more zoomed out for the lower-mass galaxies than in Fig.~\ref{fig:nHmaps}, and that they all have the same physical scale, in order to give a better impression of the difference in size between the different mass galaxies.}. With the exception of a few expanding bubbles of very hot gas originating from SN explosions close to the midplane, the three discs are dominated by gas at temperature around or below $10^5\,\rm K$ when CR transport is included. We come back to the temperature phase of CR-driven outflows in Section~\ref{subsection:cgm}.


\subsection{Regulation of star formation}
\label{subsection:sfr}


\begin{figure}
	\includegraphics[width=7cm]{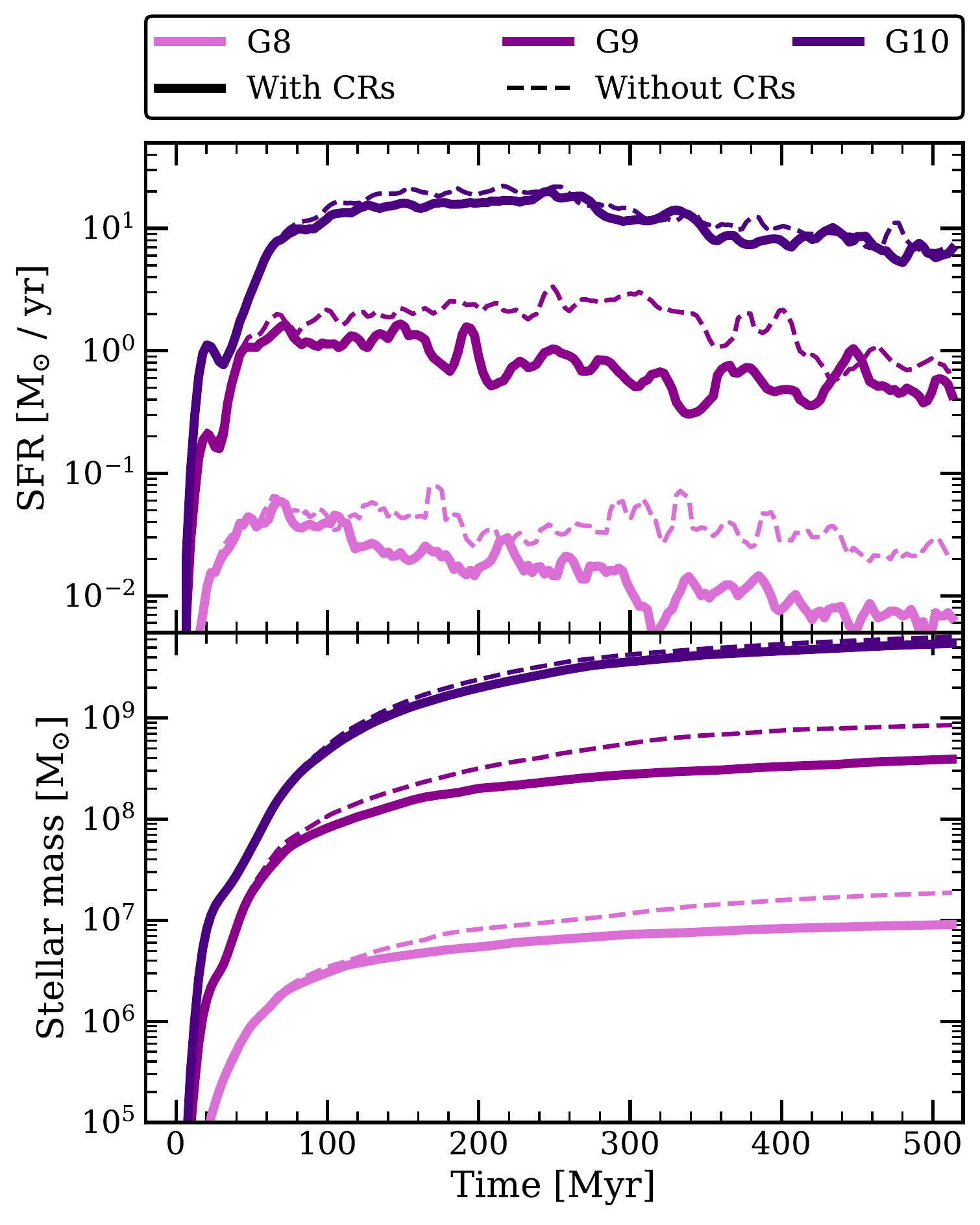}
	\centering
    \caption{Star formation rate (upper panel) and stellar mass (lower panel) versus time for G8 (light purple), G9 (purple) and G10 (dark purple). We show the runs including CRs in solid line and the runs without CRs in dashed line. We exclude the initial stellar particles seeded in the initial conditions of the discs, to show only the stellar mass formed since the start of the run. We note a reduction of the total stellar mass by a rough factor 2 for the two dwarf galaxies when we include CRs, while the star formation history of G10 does not seem affected much.}
    \label{fig:sfr}
\end{figure}

Figure~\ref{fig:sfr} shows the effect of CR feedback on star formation for our three galaxies. The upper panel shows the star formation rates (SFR, averaged over 10 Myr) and reveals a bursty star formation, particularly for the two lower mass galaxies~\citep[see also][]{Faucher2018}. CR feedback regulates the SFR for G8 and G9 after the initial collapse taking place during the first 100 Myr or so. In G10, however, the star formation is barely impacted by the CR feedback.

Globally, CRs have a significant effect on the amount of stars formed. As we can see in the lower panel of Fig.~\ref{fig:sfr}, they suppress the total star formation over the modelled 500 Myr by around a factor 2 in the lower-mass galaxies, with a decreasing efficiency with increasing mass. The same factor 2 in star formation reduction for our two dwarf galaxies is found by \DDtw{}. This is despite our different setups, where we also account for radiation feedback, non-equilibrium chemistry and a more bursty and physically motivated star formation model. This implies that the efficiency of CRs in regulating star formation does not depend strongly on the inclusion of radiative feedback or the star formation model (see also Appendix~\ref{app:dens-fk2} for a comparison of star formation history with the two star formation models). A broader discussion of our results compared to other works is provided in Section~\ref{section:discussion}.


The addition of SN feedback to the no feedback case (not shown) reduces star formation by 85, 45 and 40\% in G8, G9 and G10, respectively. In our simulations with both cosmic rays and SN feedback, we find a further suppression with respect to the SN feedback case of 50\% for the dwarf galaxies (G8 and G9), and 14\% for our most massive mass galaxy (G10). Therefore, the star formation suppression efficiency decreases with increasing galaxy mass for both SN and CR feedback.

\begin{figure}
	\includegraphics[width=\columnwidth]{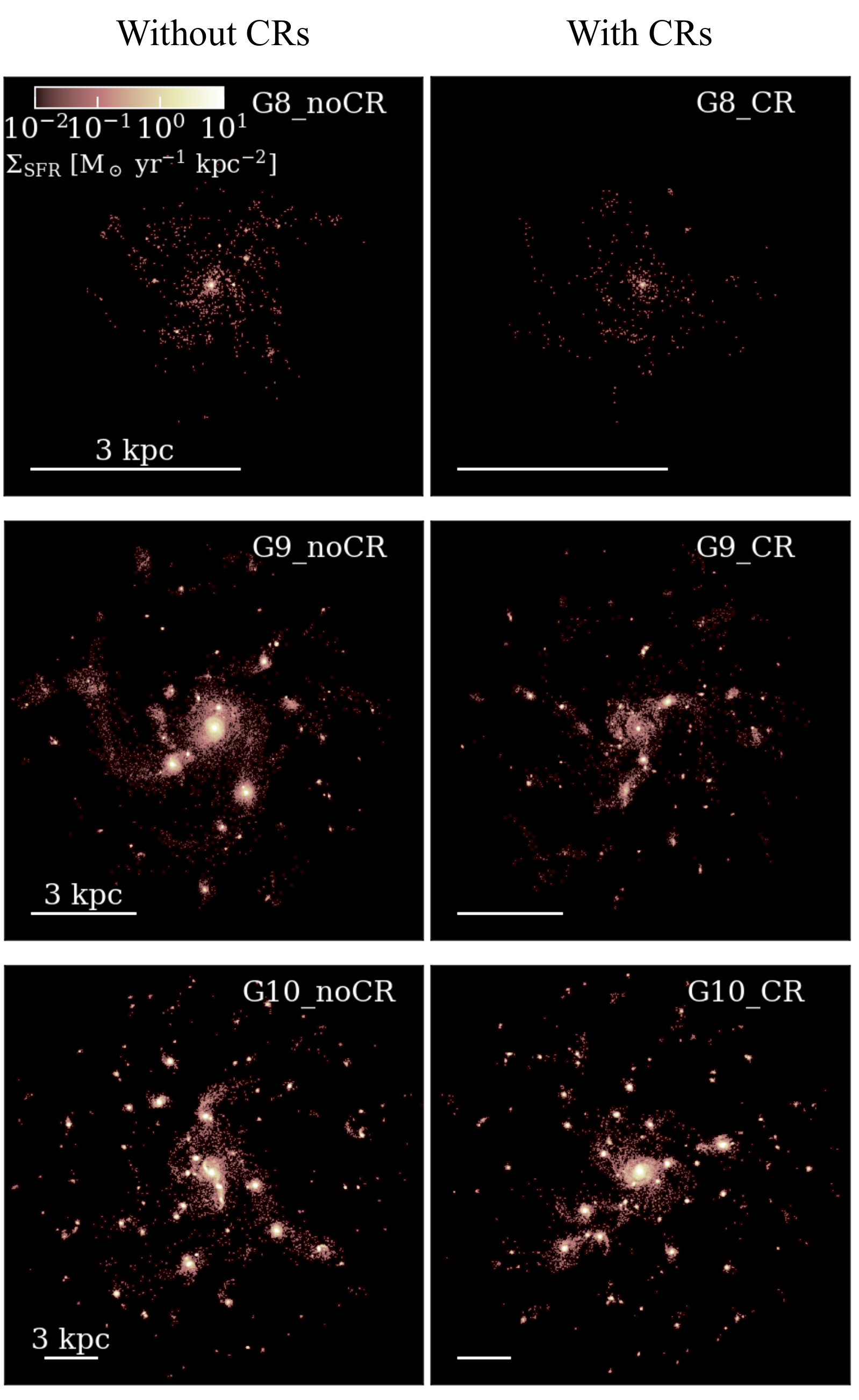}
	\centering
    \caption{Face-on maps of SFR surface density at t = 350 Myr, in order of increasing galaxy mass from the top to the bottom. Left and right columns show the simulations without and with cosmic ray feedback, respectively. The star formation rate values are derived from the last 100 Myr. The maps are decomposed in 1024x1024 squared pixels, with values smoothed by a Gaussian filter of one pixel width, for a better visibility. It is especially clear for the dwarf galaxies that adding CRs leads to less numerous and massive stellar clumps.}
    \label{fig:sfrmaps}
\end{figure}

In Fig.~\ref{fig:sfrmaps}, we show face-on maps of the SFR surface density 350 Myr after the start of the simulations, with the SFR averaged over 100 Myr. The maps reveal the ability of CRs to reduce the number and mass of stellar clumps. This is a consequence of CR feedback smoothing out the inner gas distribution of the ISM, as shown in Fig.~\ref{fig:nHmaps}. This effect is especially visible for the lower mass galaxies, where CR feedback significantly regulates the total SFR. However there is also a somewhat reduced "clumpiness" in the case of G10 where the total star formation is not diminished. 


\begin{figure}
    \centering
    \includegraphics[width=7.5cm]{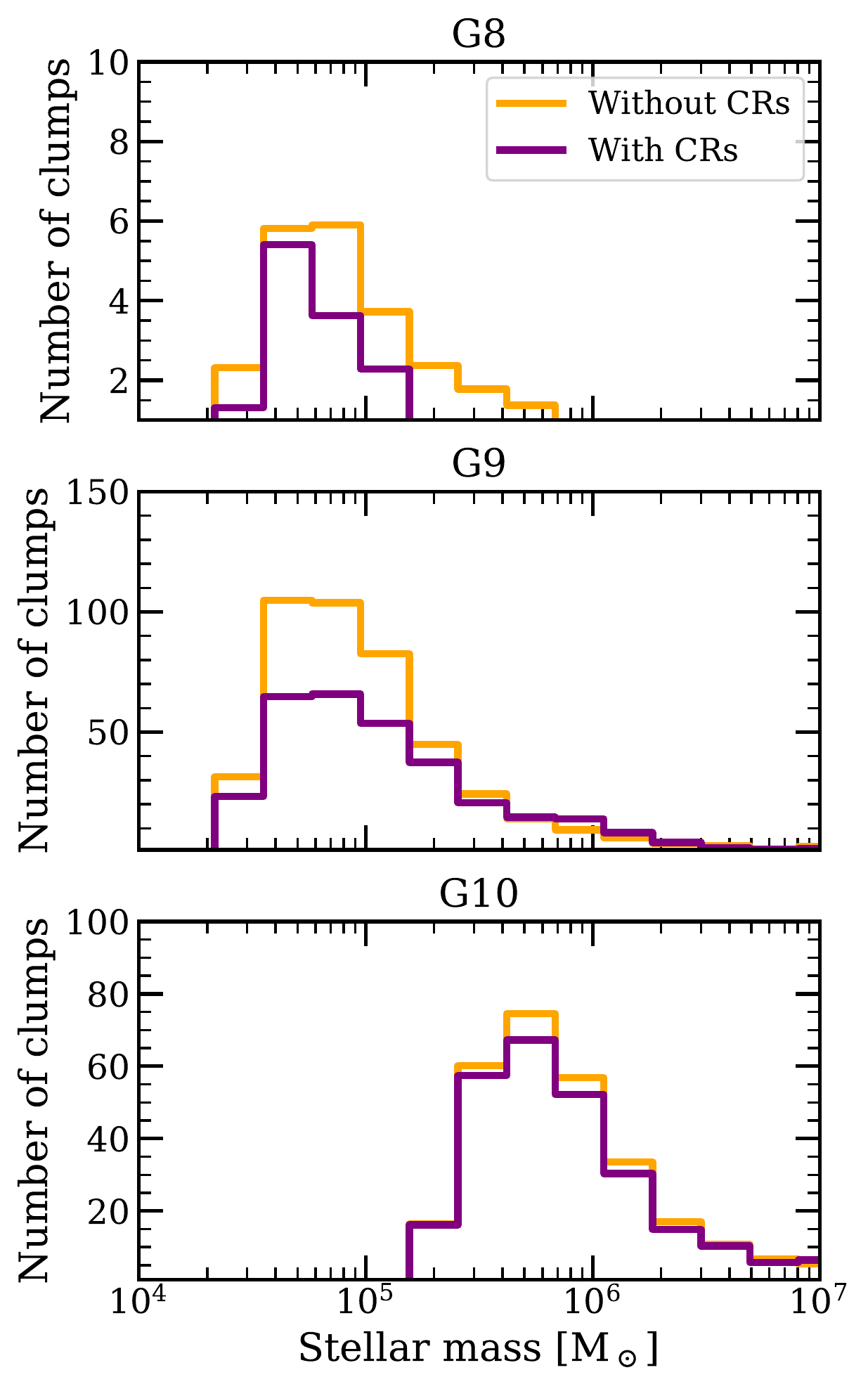}
    \caption{Number of stellar clumps as a function of their mass, with 15 logarithmic bins between $10^4$ and $10^7\ \rm M_\odot$. The panels represent increasing galaxy mass from top to bottom. Orange and purple colours correspond respectively to galaxies without and with CRs, and we show the average number of clumps in each mass-bin for outputs stacked between 200 and 500 Myr. CRs reduce the number and the mass of the stellar clumps in all our galaxies, but less efficiently with increasing galaxy mass.}
    \label{fig:nbclumps}
\end{figure}

We quantify the clumpiness at ISM scales in Fig.~\ref{fig:nbclumps}. For our three galaxies with (purple) and without (orange) CRs, we show the mass distribution of stellar clumps. The number of clumps in each mass bin is averaged by stacking data from 200 to 500 Myr, a time interval for which the SFR is roughly constant. To identify the clumps, we use the \adaptahop{} algorithm in the most massive substructure mode \citep{Aubert2004,Tweed2009}. Following the notation used in \citet[][in Appendix B]{Aubert2004}, we adopt $N_{\rm SPH}~=~16$, $N_{\rm HOP}~=~8$, $\rho_{\rm TH}~=~80$ and $f_{\rm Poisson}~=~2$. Then, we define a clump as the closest stellar particles (at least 10) to a common local maximum, corresponding to the centre of the clump.

When CR feedback is included, the number of stellar clumps is strongly suppressed in the two lower-mass galaxies, as also visible in Fig.~\ref{fig:sfrmaps}. However, CRs only marginally reduce the number of clumps in G10. We additionally note that there are fewer clumps at low masses in G10, compared to what is measured for G8 and G9. This is due to the coarser resolution in G10, which has stellar particles at least 8 times more massive than our dwarf galaxies (see Table~\ref{tab:run_prop}). Because of the lower limit in the number of particles per clump set when using the \adaptahop{} algorithm, the lower mass of a stellar clump in G10 is higher than that in our two dwarf galaxies.


\begin{figure}
    \centering
    \includegraphics[width=7.5cm]{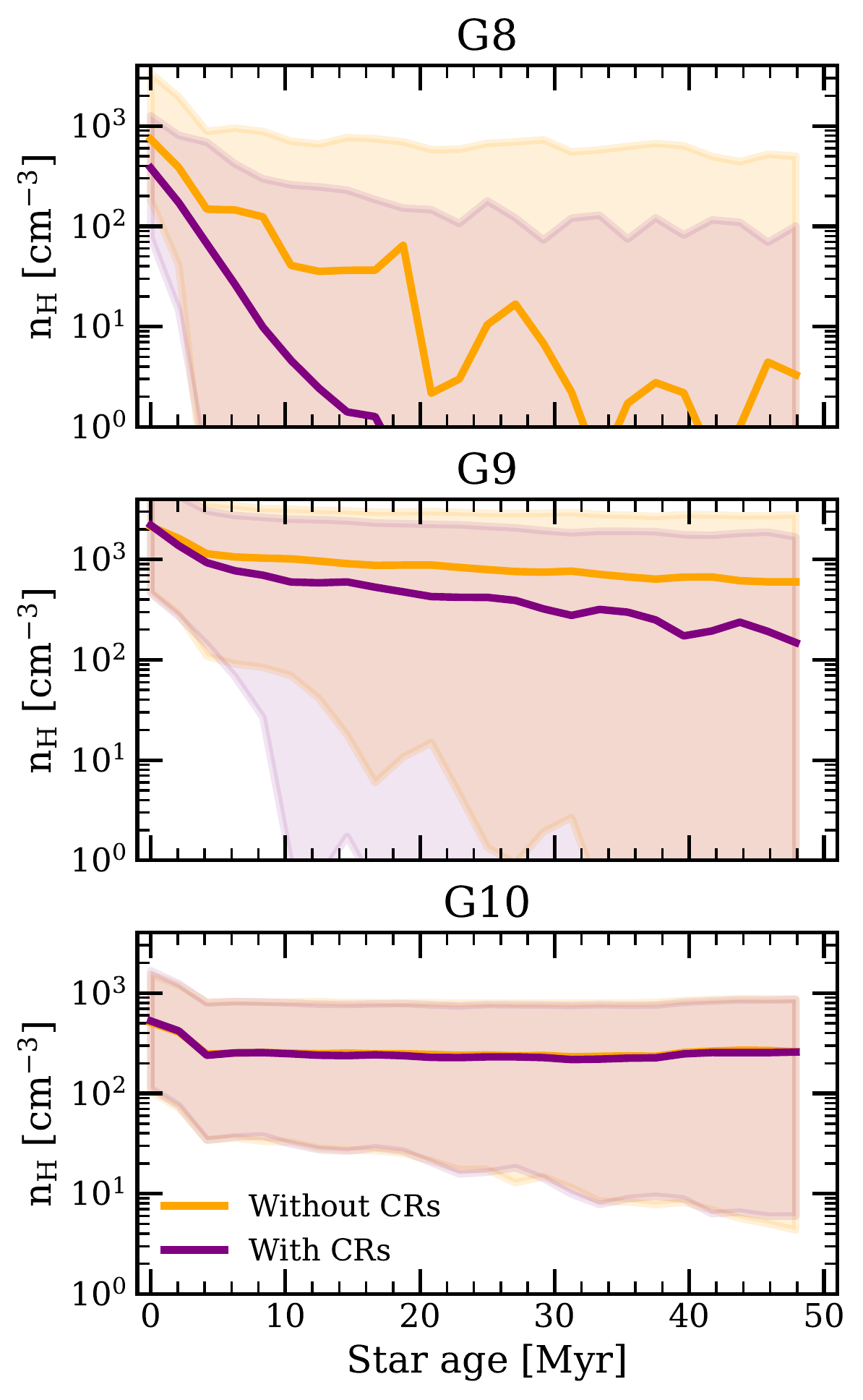}
    \caption{Densities of cells hosting stellar particles as a function of the particle age, binned every 2 Myr. The panels represent increasing galaxy mass from top to bottom. Solid lines show the median density in each age-bin for outputs stacked between 200 and 500 Myr. The shaded areas give the 10th and 90th percentiles in each stellar age bin. Orange and purple colours correspond respectively to galaxies without and with CRs. With increasing galaxy mass, CRs become less efficient in dispersing gas around the sites of star formation.}
    \label{fig:nstars}
\end{figure}

In order to explain the reduction of stellar clumps with CRs, Fig.~\ref{fig:nstars} explores the efficiency of CR feedback in dispersing gas locally at the sites of star formation. For each galaxy, the histograms show the median density of the cells in which the stellar particles are located as a function of their age, binned every 2 Myr. To avoid any transient effect, the density of each bin is averaged from stacking the outputs between 200 and 500 Myr in steps of 10 Myr. We show results from the runs with (without) CRs in purple (orange).

Each stellar particle undergoes several SN explosions between 3 and 50 Myr, which disperse gas locally and reduce the local density as the particles age. When CRs are injected from these SN explosions, they further disperse local densities around young stars in our dwarf galaxies. Because G8 has a shallow gravitational potential, the gas dispersal caused by the CR pressure has more visible consequences than for our two other galaxies. After 20 Myr, only half of its stellar particles are surrounded by gas more diffuse than a few atoms per $\rm cm^3$ when CRs are included. Not only are stellar clumps rapidly dispersed, but star formation also occurs at slightly lower densities with CRs added, which is not the case for the two other galaxies. Nonetheless, the density of gas in star-forming regions decreases for both G8 and G9 after the first SN explosions, especially when CRs are injected, which leads to the reduced number and masses of stellar clumps shown in Figures~\ref{fig:sfrmaps} and \ref{fig:nbclumps}. In G10 however, CRs do not act strongly enough to disperse gas, and the density of gas in star-forming regions remains the same with and without CRs. Similarly, photoionization heating from young stars has been shown to have a similar effect on local gas densities at G8 and G9 galaxy masses, but also to have negligible effect in G10 \citep{Rosdahl2015}. We additionally note that stars form at lower densities in G10 than in G9 due to the coarser resolution in G10, which does not affect the strength of CR feedback (see Section~\ref{section:discussion}).

We summarise the effects of CRs on star formation as follows. In low-mass galaxies, they disperse gas in the vicinity of the SNe away from star-forming clumps, which smooths the ISM. CRs therefore delay gas in reaching the density needed to form stars, which suppresses star formation, as less numerous and/or less massive clumps can form.

\subsection{Outflows}
\label{subsection:cgm}

\begin{figure*}
    \centering
	\includegraphics[width=\textwidth]{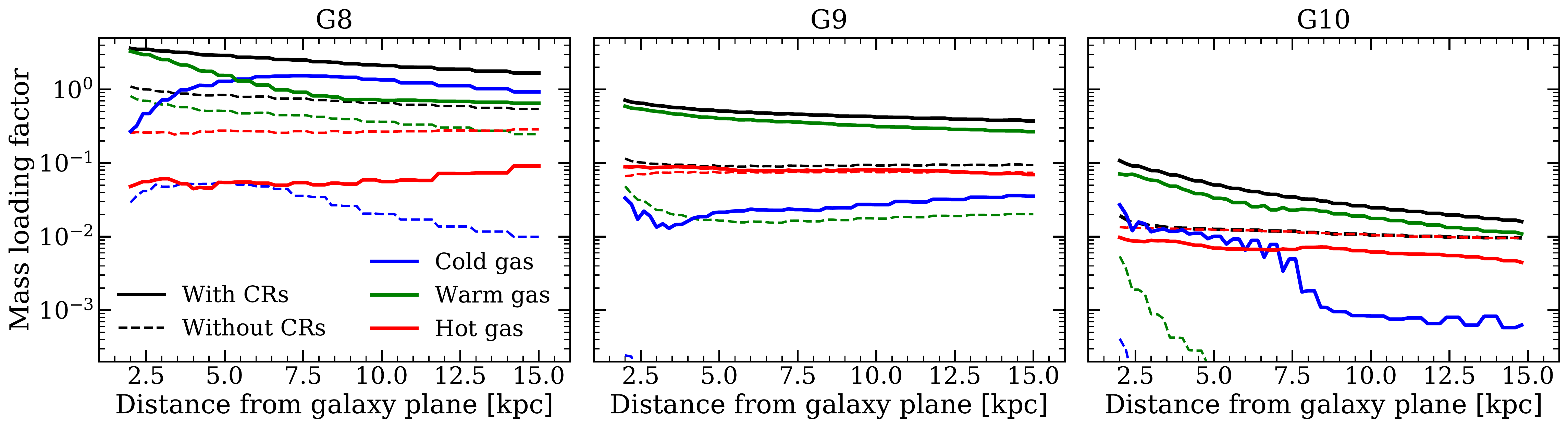}
    \caption{Mass loading factors (ratio of outflow rate to star formation rate) of gas crossing slabs at different distances from the galaxy midplane for G8 (left), G9 (middle) and G10 (right), with data stacked from 200 to 500 Myr. The blue, green and red lines respectively stand for cold (T $< 10^4$ K), warm ($10^4$ K $\le$ T $< 10^5$ K) and hot (T $\ge 10^5$ K) gas. We show the mass loading factors for the total gas without any temperature distinction in black. The dashed lines are for the runs without CRs, where no (or almost no) cold gas is outflowing at any time and no matter the galaxy mass, and the solid lines are for the runs with CRs added, with more outflowing gas in total, dominated by a warmer phase and with more cold gas than the noCR run counterparts.}
    \label{fig:eta}
\end{figure*}


One of our main goals is to assess the role of CR feedback in launching gas from galaxy discs, and especially its ability to push away cold material. To quantify the efficiency of feedback in generating galactic winds, we focus on the mass loading factor, defined as the mass outflow rate normalised by the star formation rate. By outflowing gas, we mean all gas which is flowing away from the disc in the vertical direction. In order to avoid spurious oscillations in the mass loading factor due to the bursty star formation and the delay between starbursts and an increase in outflows kiloparsecs away from the disc, we use star formation rates averaged over the last 50 Myr. To measure the outflow rate, we define planes parallel to the disc at a given distance from it. For each cell, the rate of outflowing gas mass $\dot{m}_{\rm cell}$ is defined as the product of the gas density ($\rho_{\rm cell}$) with its vertical velocity ($u_{z,\rm{cell}}$) and the surface of the cell ($\Sigma_{\rm cell}$), i.e.: $\dot{m}_{\rm cell}=\rho_{\rm cell}u_{z,\rm{cell}}\Sigma_{\rm cell}$. The total mass outflow rate is then derived by summing the values of all the cells intersected by the selected planes.

To better study the impact of CRs on the outflowing gas phase, we distinguish three temperature regimes, namely cold for gas with $T < 10^4\ K$, warm for $ 10^4 \leq T < 10^5$ K and hot for gas at temperature~$\geq 10^5$ K. These temperature ranges are chosen to trace observational lines. What we call cold mainly corresponds to neutral gas, the warm phase can be traced through MgII, CIII or SiIV absorption lines, and the hot gas can be detected with X-ray emission, CIV or OVI absorption lines (as has been done in the COS-haloes survey data from \citealp{Werk2013, Werk2016}).

Fig.~\ref{fig:eta} shows profiles of the mass loading factor as a function of distance from the disc plane. In order to reduce the noise due to transient effects and bursty star formation, we stack 32 outputs between 200 and 500 Myr. Each panel contains what we define to be cold, warm or hot gas both for the galaxies without (in dashed lines) and with (in solid lines) CR feedback.

Focusing first on the black solid and dashed lines, adding CR feedback leads to a net increase in the loading factor at all masses. The same behaviour was qualitatively found by \DDtw{}, but with an even stronger effect from CRs on driving winds. We detail the reasons for this difference in Section~\ref{section:discussion}.

Without CRs, the outflow is dominated by the hot phase in all three galaxies, with only a small amount of warm component and a tiny fraction of cold gas ejected. In contrast, with CRs, the outflows become preferentially warm, and cold outflowing gas can be found at any distance from the disc, even if in a smaller proportion for the two more massive galaxies. Measurements of MgII absorption in quasar sightlines around galaxies appear to disfavour the complete lack of warm gas produced in our non-CR simulations \citep{Bordoloi2011,Bouche2012}. We will study more quantitatively the effect of CR feedback on the MgII content around galaxies and compare to observations in upcoming work.

We note that the less massive the galaxy, the higher the mass loading factor, independently of the feedback \citep[in agreement with observations, e.g.][]{Heckman2015}. Because dwarf galaxies have a shallower potential well, we can expect that stellar feedback can expel gas more efficiently~\citep{Dubois2008}. We also find a trend of decreasing outflow rate with distance, most particularly for G10 when CRs are added. For the latter, the amount of cold gas in the outflows suddenly drops, especially above 7.5 kpc. 

\subsection{Sensitivity to the cosmic ray diffusion coefficient}
\label{subsection:G10_dcr}

\begin{figure}
	\includegraphics[width=\columnwidth]{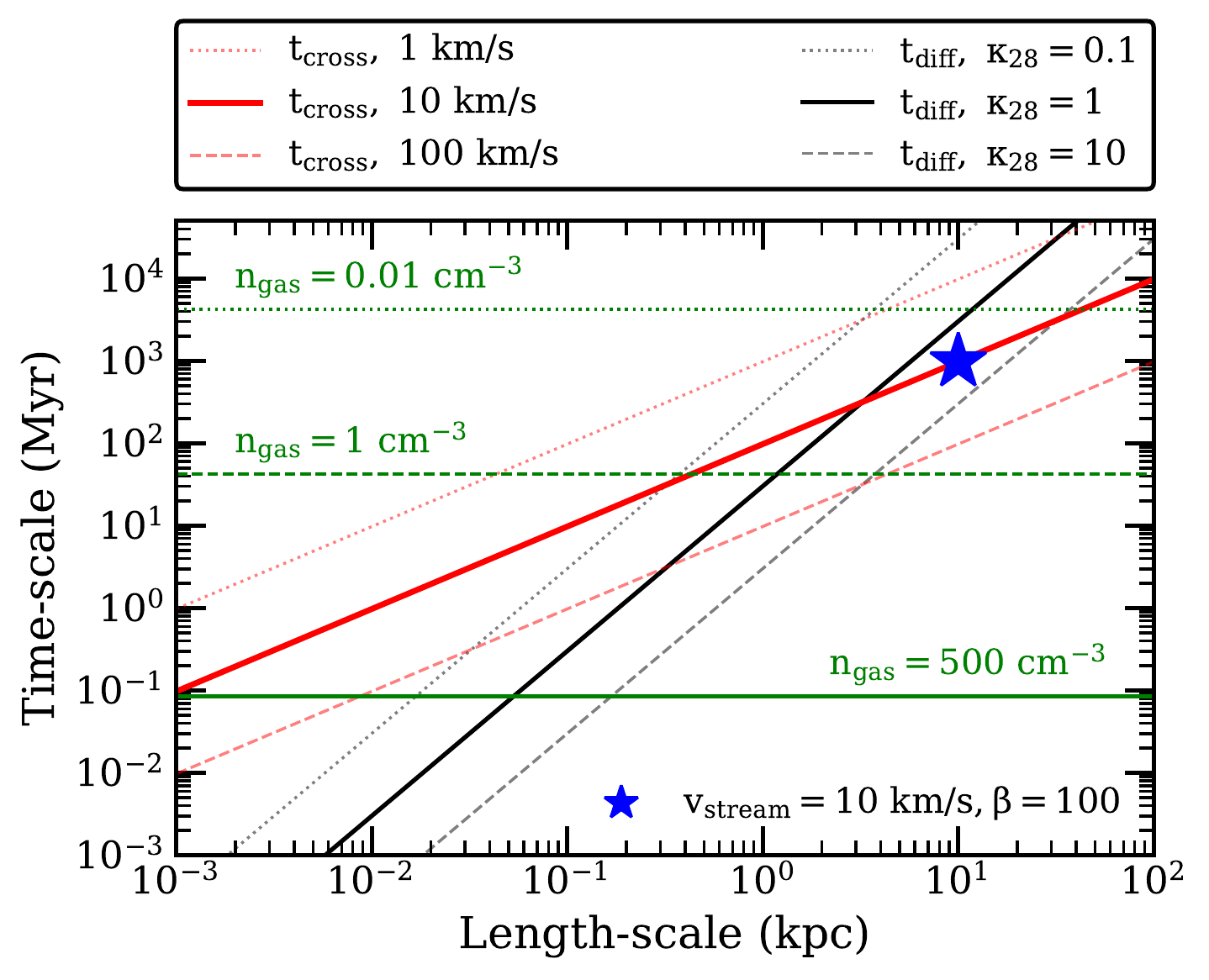}
	\centering
    \caption{Time-scale against length-scale for the crossing time (in red), CR diffusion time (in black) and CR energy dissipation time (in green) for different gas velocities, diffusion coefficients, and gas densities, as indicated in the legend. At small scales ($L \leq 1$ kpc), CR diffusion is the dominant transport process. However, the lower the diffusion coefficient, the slower the diffusion, so the more significant the CR energy losses before they are propagated to disc scales. At CGM scales ($L \sim$ 10 kpc), considering a gas velocity of $\rm \sim 100\ km\,s^{-1}$, CRs are mostly advected with gas, as CR diffusion is slower for any diffusion coefficient. At CGM gas densities ($n_{\rm H} \rm < 0.01\ cm^{-3}$), CR energy losses become negligible, and the time associated to streaming (represented by a blue star) becomes comparable to or somewhat shorter than diffusion, but remains longer than advection.}
    \label{fig:timescales}
\end{figure}

The impact of CRs, both at ISM and CGM scales, is predominantly determined by the force they apply on gas. This force directly depends on the CR pressure gradient, which evolves due to CR diffusion and dissipation, and is therefore largely ruled by their diffusion coefficient, which is a key parameter governing their propagation. We now investigate how the diffusion coefficient affects CR feedback. 

Observationally or theoretically, there are not yet strong constraints on the diffusion coefficient. Empirically, and from fitting models of CR propagation (with codes like \galprop{}, \citealp{Strong&Moskalenko1998}), we expect a diffusion coefficient of a few $10^{28}\,\rm cm^2\,s^{-1}$. In addition, the diffusion coefficient is not homogeneous but rather depends on the energy of the CR particles \citep{Zweibel2013}, as well as on the local gas properties, such as the level of turbulence and the ionisation fraction \citep[e.g.][]{Bustard&Zweibel2021}. For simplicity and computational efficiency, simulations that include CRs generally adopt a constant diffusion coefficient, with values typically varying from $10^{27}$ to a few $10^{29}\,\rm cm^2\,s^{-1}$ from one work to another (see for instance \citealp{Salem2016, Pakmor2016,Girichidis2018, Farber2018, Buck2020,Dashyan&Dubois2020,Ji2020,Hopkins2020}, but also \citealp{Farber2018, Hopkins2021,Girichidis2021,Semenov2021} for a diffusion coefficient varying with gas properties or CR energy). We therefore test the variability of CR feedback by performing additional simulations with the following values: $\kappa_{28}=\left\{0.1,1,3,10,30\right\}$ where $\kappa_{28}$ is the diffusion coefficient in units of $\rm 10^{28}\ cm^2\, s^{-1}$. As \DDtw{} did a similar revision for the two lower mass discs G8 and G9, we also comment on how the results are affected by our additional physics, namely the inclusion of radiative transfer as well as more physically motivated models for star formation and SN feedback.

\begin{figure}
	\includegraphics[width=\columnwidth]{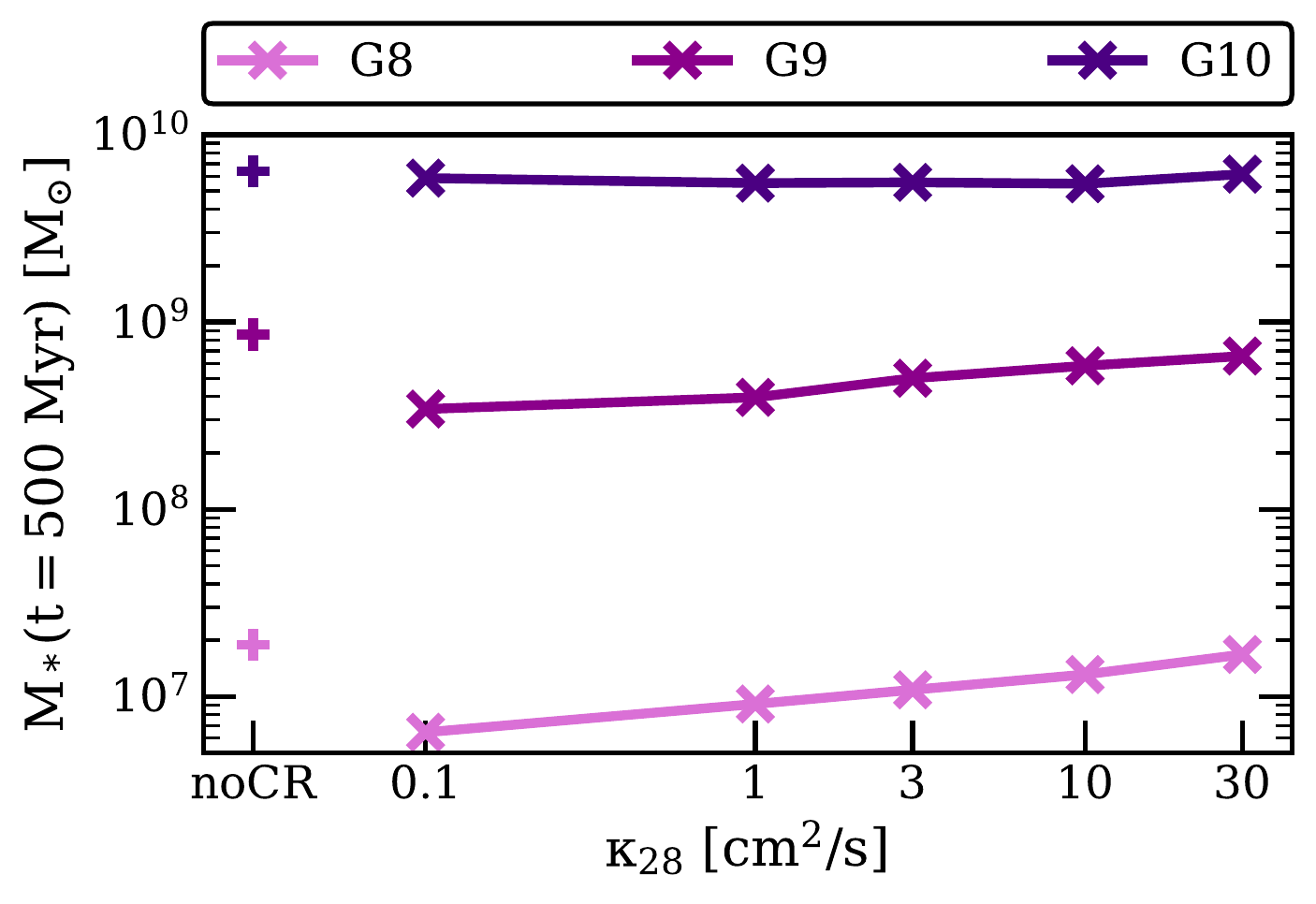}
	\centering
    \caption{Stellar mass formed by the end of the 500 Myr runtime for G8, G9 and G10 with increasing diffusion coefficient from left to right. The leftmost data point for each galaxy represents the stellar mass formed without CR feedback. Star formation is most efficiently regulated with the lowest diffusion coefficient considered. In the most massive galaxy, star formation is insensitive to CR feedback at any $\kappa$ (nor in fact is it sensitive to any feedback we include).}
    \label{fig:DCR_sm}
\end{figure}


\begin{figure*}
    \centering
	\includegraphics[width=\textwidth]{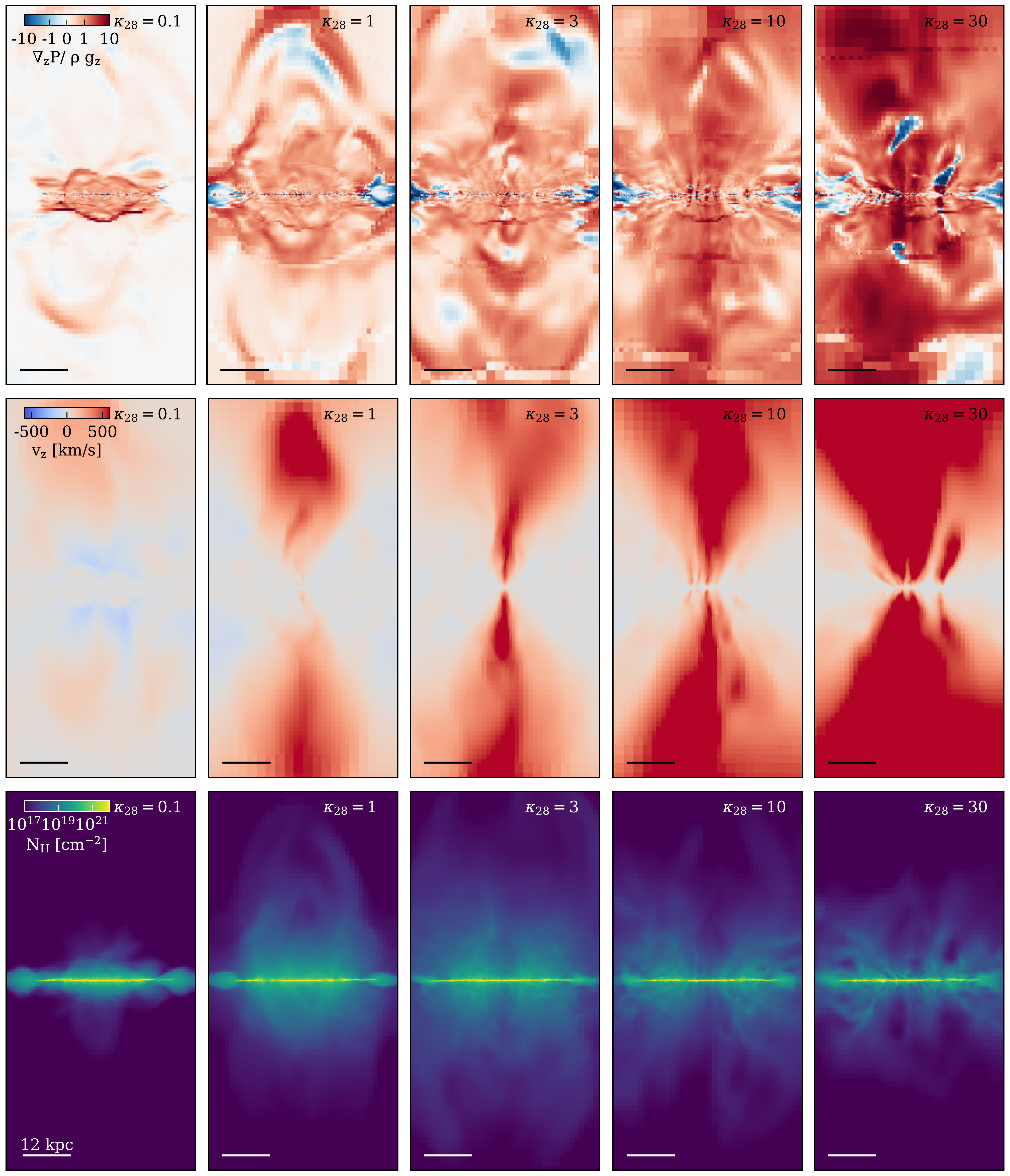}
    \caption{6 kpc-deep projections centered on G10 at 500 Myr in order of increasing $\kappa$ from left to right. For each run, we show edge-on maps of the CR pressure gradient over the vertical gravitational force (top row), vertical velocity (middle row) and hydrogen gas column density (bottom row). With increasing diffusion coefficient, the CR pressure overcomes gravity more easily. As winds are pushed faster and to larger distances, the gas distribution around the galaxy becomes increasingly extended but also more diffuse.}
    \label{fig:DCRmaps}
\end{figure*}

\begin{figure*}
    \centering
	\includegraphics[width=\textwidth]{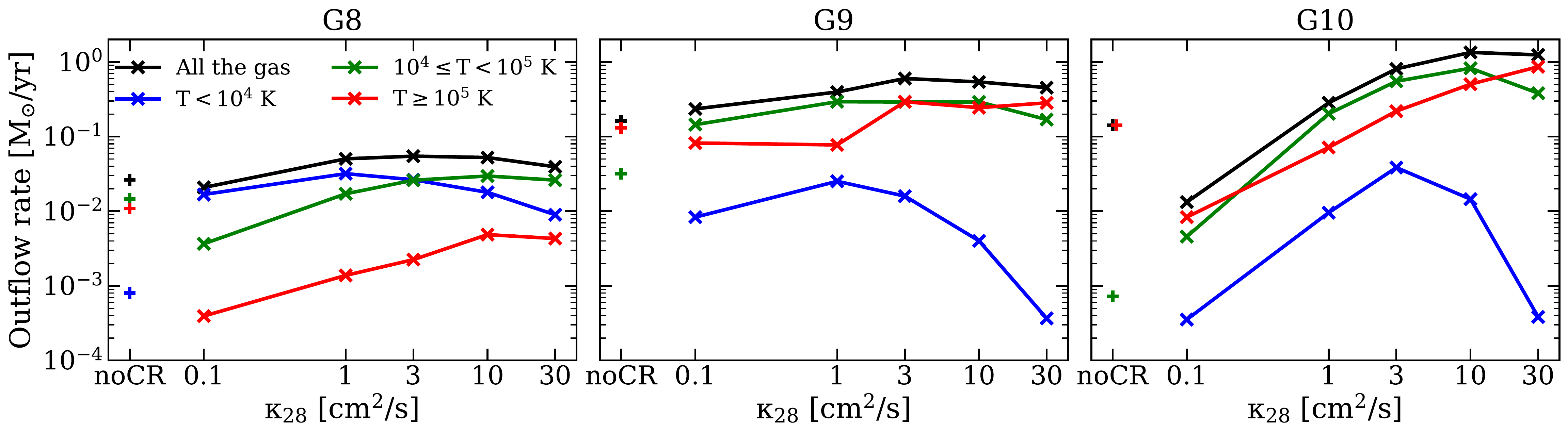}
    \caption{Mass outflow rate of gas crossing slabs at 10 kpc from the galaxy midplane as a function of the diffusion coefficient, in order of increasing galaxy mass from left to right. For each galaxy, data are stacked between 200 and 500 Myr. We show the total amount of outflowing gas in black, the cold component ($\rm T < 10^4\ K$) in blue, the warm ($\rm 10^4 \leq T < 10^5\ K$) in green and the hot ($\rm T \geq 10^5\ K$) in red. The leftmost points are for runs without CRs. The total amount of outflowing gas and especially its hot component are globally enhanced with higher values of $\kappa$. However, the rate of cool outflows stops increasing and even drops beyond a diffusion coefficient limit, which increases with galaxy mass.}
    \label{fig:outflows-DCR}
\end{figure*}

In order to appreciate the impact of CRs in our simulations, it is useful to compare time and length scales over which different competing factors operate. Being charged particles, CRs diffuse along magnetic field lines by scattering off magnetic field inhomogeneities. The characteristic time for CRs to diffuse over a length-scale $L$ with a diffusion coefficient $\kappa$ is $t_{\rm diff}=L^2/\kappa$. Because the thermal and CR components are tightly coupled to the magnetic field, CRs are also advected with the gas at the gas velocity $u$. The density of CRs thus evolves on a timescale related to a crossing time $t_{\rm cross} = L/u$. CRs can also be transported by streaming, which occurs along magnetic field lines and down the CR pressure gradient at about the Alfvén speed $u_{\rm A}$\footnote{The exact speed depends on the major damping process of the CR-excited Alfvén waves}. Streaming has been shown e.g. by \DDtw{} to be a subdominant process compared to advection and diffusion. Therefore, we do not expect that streaming affects our results, and we do not include it in our simulations. Finally, CRs dissipate energy at a rate which scales with the gas density $n_{\rm gas}$. For Coulomb and hadronic collisions, CRs lose energy at a rate $\Gamma_{\rm CR} = \xi_{\rm coll} \times (n_{\rm gas}/{\rm cm^{-3}})\times(e_{\rm CR}/{\rm erg\ cm^{-3}})\ {\rm erg\ s^{-1}\ cm^{-3}}$, where $e_{\rm CR}$ is the CR energy density and $\xi_{\rm coll} = 7.51\times 10^{-16}\ {\rm cm^3\ s^{-1}}$ is the rate of collisional CR energy loss (\citealp{Guo&Oh2008}, and as implemented in the \ramses{} code used in this study. See also equations~\ref{eq:etot} and \ref{eq:ecr}). The corresponding CR energy loss time-scale is therefore $t_{\rm loss} = (\xi_{\rm coll}\times n_{\rm gas})^{-1}$.

We illustrate these scaling behaviours for parameter values of interest on Figure~\ref{fig:timescales}, which shows transport time-scales in Myr against transport length-scales in kpc. The time-scales associated to CR diffusion are shown with black lines. The crossing time associated to CR advection is plotted with red lines. Green horizontal lines indicate the CR energy dissipation time-scales at the corresponding gas density. We use Fig.~\ref{fig:timescales} in the two following subsections to analyse why different diffusion coefficients lead to different consequences on star formation and launching of winds.

\subsubsection{Cosmic rays in star-forming clouds}

We first focus on molecular cloud scales of around 50 pc, as this is where CRs are injected when SNe explode. At these scales, the typical gas velocity is $1-10 \,\rm km\,s^{-1}$, corresponding to the dotted and solid red lines in Fig~\ref{fig:timescales}. If we compare them to the black solid line, which corresponds to $\kappa_{28}=1$, we see that the diffusion time is shorter than the crossing time. This is the case at any diffusion coefficient in the range of values we show in the plot, meaning that typically the diffusion of CRs is much faster than their advection with gas at small scales. The escape of CRs from star-forming regions is therefore ruled by diffusion, and thus by the diffusion coefficient. With lower diffusion coefficient, CRs are stuck for longer in the ISM, so they have more time to disrupt star forming clouds. Consequently, we expect CRs to be more efficient at suppressing star formation with a low diffusion coefficient.

Figure~\ref{fig:DCR_sm} compares the stellar mass formed during the 500 Myr runtime for our three discs, with increasing diffusion coefficient. For the two dwarf galaxies (G8 and G9), we find the largest regulation of star formation when the diffusion coefficient is the lowest, as found by \citet{Salem&Bryan2014}, \citet{Chan2019} and \DDtw{}, and as expected from our length versus time-scale analysis from Figure~\ref{fig:timescales}. As lower diffusion coefficient leads to a slower CR diffusion, it also leads to a stronger direct effect of CRs on the star-forming regions and, hence, on star formation. For G10 however, the effect of CRs on star formation remains weak at any $\kappa$, as do the other forms of (SN and radiation) feedback modeled here \citep[see][]{Rosdahl2015}.

Another important process in dense star-forming clouds is the dissipation of CR energy. If embedded in a $\rm 500\, cm^{-3}$ density gas, CRs lose their energy over time-scale of about 0.04 Myr, as shown by the solid green line in Fig~\ref{fig:timescales}. This energy dissipation time is shorter that the diffusion time-scales for $\kappa_{28} \leq 1$. As a result, CRs with low $\kappa$ lose a large amount of their energy before they can reach less dense gas on larger scales, and we expect a lower impact from CRs at large scales in this case. We now assess whether this is indeed the case.

\subsubsection{Cosmic rays in the CGM}

Before assessing the effects of changing the CR diffusion coefficient on CGM gas, we first come back to our length versus time-scale analysis from Figure~\ref{fig:timescales}. At nearby CGM scales, the crossing time corresponds to a gas velocity of $100 \,\rm km\,s^{-1}$, typical for both the sound speed and outflow velocities in the CGM. As it becomes smaller than the diffusion time, this shows that CRs are mainly advected with gas rather than via diffusion. In this medium, the gas has densities lower than $\rm 0.01\, cm^{-3}$, for which the loss of CR energy occurs on time-scales of several Gyr and hence is completely subdominant. The CR energy is thus conserved and propagates through the CGM.

It is only at CGM scales that CR streaming is faster than diffusion. For this reason, we show with a blue star the time associated to streaming at a scale of 10 kpc, where the plasma $\beta$ (ratio of thermal to magnetic pressure) is around 100. As $v_{\rm A} \propto c_{\rm s}/\sqrt{\beta}$, at 10 kpc where the sound speed $c_{\rm s} \simeq 100\,\rm  km\,s^{-1}$, $v_{\rm stream}\simeq 10\,\rm  km\,s^{-1}$. Comparing the blue star to the red dashed line, we can see that the streaming time-scale is around $10^3$ Myr, which is ten times longer than the advection. Even at CGM scales, streaming is subdominant compared to the transport of CRs via gas advection. To show a significant contribution, the streaming velocity has to be boosted by damping effects, such as ion-neutral damping or turbulence as in \citet{Ruszkowski2017} and \citet{Hopkins2021}. Even so, these two studies have opposite conclusions on the importance of the role of CR streaming, which remains a topic of extensive investigation.

To summarize, Figure~\ref{fig:timescales} shows that diffusion is the dominant process in the CR injection sites, meaning that the diffusion coefficient directly impacts the confinement of CRs in dense regions of the galaxy, where most of their radiative energy losses occur. The competition of diffusion and CR energy losses in dense gas regulates the amount of CR energy escaping into the more diffuse ISM and hence, potentially, their impact on larger scales. We will now assess whether varying the diffusion coefficient in our simulations has the effects predicted by these scale-comparisons.

Varying the diffusion coefficient has a strong effect on gas morphologies and outflows in all our galaxies. Figure~\ref{fig:DCRmaps} qualitatively illustrates the effect of the diffusion coefficient on G10, with increasing $\kappa$ from left to right. The top row shows the ratio of the CR pressure gradient (i.e. the force from the CR pressure) to the vertical gravitational force of the disc, where red (blue) cells have an outward (inward) net force. Red colours thus show cells where the force exerted by CRs can overcome the gravitational potential of the galaxy. The middle row shows the vertical velocity of the gas, with red (blue) colours for outflowing (inflowing) gas. The bottom row finally shows the hydrogen column density.

The force exerted by CRs strongly affects the gas distribution around the disc. The $\rm \kappa=10^{27}\,\rm cm^2\,s^{-1}$ case merely produces a closely confined outflow fountain, as the CR gradient vanishes beyond a few kpc, which leads to a thick disc surrounded by dense gas. When we increase the diffusion coefficient, we find larger CR pressure gradients at larger distances, producing strong bipolar winds. The gas is expelled farther away from the ISM and at higher speed, and is more broadly distributed around the disc, becoming very diffuse for the highest values. Qualitatively, the same holds for our lower mass galaxies (not shown).

Figure~\ref{fig:outflows-DCR} shows the effect of $\kappa$ on the mass outflow rate for the different galaxies, in terms of total outflowing gas and its cold ($T<10^4$ K), warm ($10^4\leq T<10^5$ K) and hot ($T\geq10^5$ K) components measured at 10 kpc from the discs. For each simulation we take the average from 31 snapshots (with 10 Myr intervals) between 200 and 500 Myr. The leftmost symbols in each plot represent runs without CR feedback.

For all three galaxies, increasing the diffusion coefficient leads to stronger and eventually hotter outflows, with total outflow rates that increase, top out and finally stagnate. Our time-scale analysis at ISM and CGM scales from Figure~\ref{fig:timescales} indicates that increasing the diffusion coefficient leads to more efficient escape of CRs from dense regions in the galactic disc, meaning less radiative losses and so more energy available to push and maintain outflows at high velocities (see also Fig.~\ref{fig:DCRmaps}). This explains the increasing outflow rates with increasing $\kappa$. This also explains why the maximum of the outflow rate with $\kappa$ does not correspond to the maximum star formation regulation. For efficient regulation of star formation, CRs have to be trapped in clouds to build a strong CR pressure gradient \citep{Commercon2019}, whereas launching winds requires CRs to escape these dense regions, which is an opposite condition to the regulation of SF by CRs. 

On long enough time-scales, the galactic winds may lead to gas-depletion in the disc, which in turn would lead to a regulation of star formation. However such timescales are beyond our simulation runtimes. We estimate depletion timescales of approximately 1.7 (G8), 1.7 (G9), and 17 Gyr (G10), where we have assumed constant outflow rates of 0.1, 1, and 1 $\rm M_\odot / yr$, respectively. For G10 in particular, this outflow depletion timescale is significantly longer than the star formation depletion time. Therefore, outflows are not expected to have a significant effect on star formation. For our lower mass galaxies, the relevance of such sustained outflows remains unclear, and we will review this with cosmological simulations in upcoming work.

In agreement with our results, \DDtw{} found for their G8 and G9 counterparts that the higher the diffusion coefficient, the stronger the outflows (see also Section~\ref{section:discussion} for more details). However, we additionally report that the outflow rate does not increase steadily with $\kappa$. Furthermore, the warm and cold outflow rates peak and then drops beyond a certain $\kappa$, which depends on the galaxy mass. 

This trend is especially strong for the cold outflowing gas. For the two dwarf galaxies (G8 and G9), the fraction of cold outflows peaks for $\rm \kappa=10^{28}\,\rm  cm^2\,s^{-1}$, but it peaks at $\rm \kappa=3\times10^{28}\,\rm cm^2\,s^{-1}$ for G10. This hints towards the existence of a diffusion coefficient value beyond which the effect of CRs gradually vanishes, dependent on galaxy mass, or alternatively its size. The more massive the galaxy, the thicker the galactic disc CRs have to cross before propagating to the CGM. For more massive galaxies, with larger length scales, a higher diffusion coefficient is then needed for CRs to escape dense regions (as shown in Fig.~\ref{fig:timescales}) and drive winds. This explains why the ability of CRs to drive winds starts vanishing at a diffusion coefficient higher for larger galaxies.

When the diffusion coefficient is high enough for CRs to quickly escape from the disc ($1-10$ Myr, see Fig.~\ref{fig:timescales}), the CR pressure starts acting at larger distance from the midplane, where gas is more diffuse, and CRs do not impact the densest and coolest gas of the galaxy anymore. Because the CR pressure gradient builds up farther away with increasing $\kappa$, the density of cold and warm outflows decreases drastically, explaining why we measure less outflowing gas at temperature below $10^5\,\rm K$.

It is also interesting to note the $\rm \kappa=10^{27}\,\rm  cm^2\, s^{-1}$ case for G10, which has smaller outflow rates at 10 kpc than its counterpart run without CRs. This is the consequence of CRs acting locally and puffing up the galactic disc, carrying with them high density gas but at velocities too low to escape the gravitational potential of the galaxy. There are slightly stronger outflows with $\rm \kappa=10^{27}\,\rm  cm^2\, s^{-1}$ than without CRs in G10 at 2 kpc (see Fig.~\ref{fig:outflows-DCR2}), but the dense outflowing material quickly falls back to the galaxy and the outflows are not maintained at large distances. When the diffusion coefficient is very small, CRs become irrelevant in driving significant outflows, and can actually become counter-productive in driving galactic winds on large scales.
\subsection{Do cosmic rays provide the needed feedback in high-redshift galaxies?}
\label{subsection:boostvscr}

\begin{figure*}
    \centering
	\includegraphics[width=\textwidth]{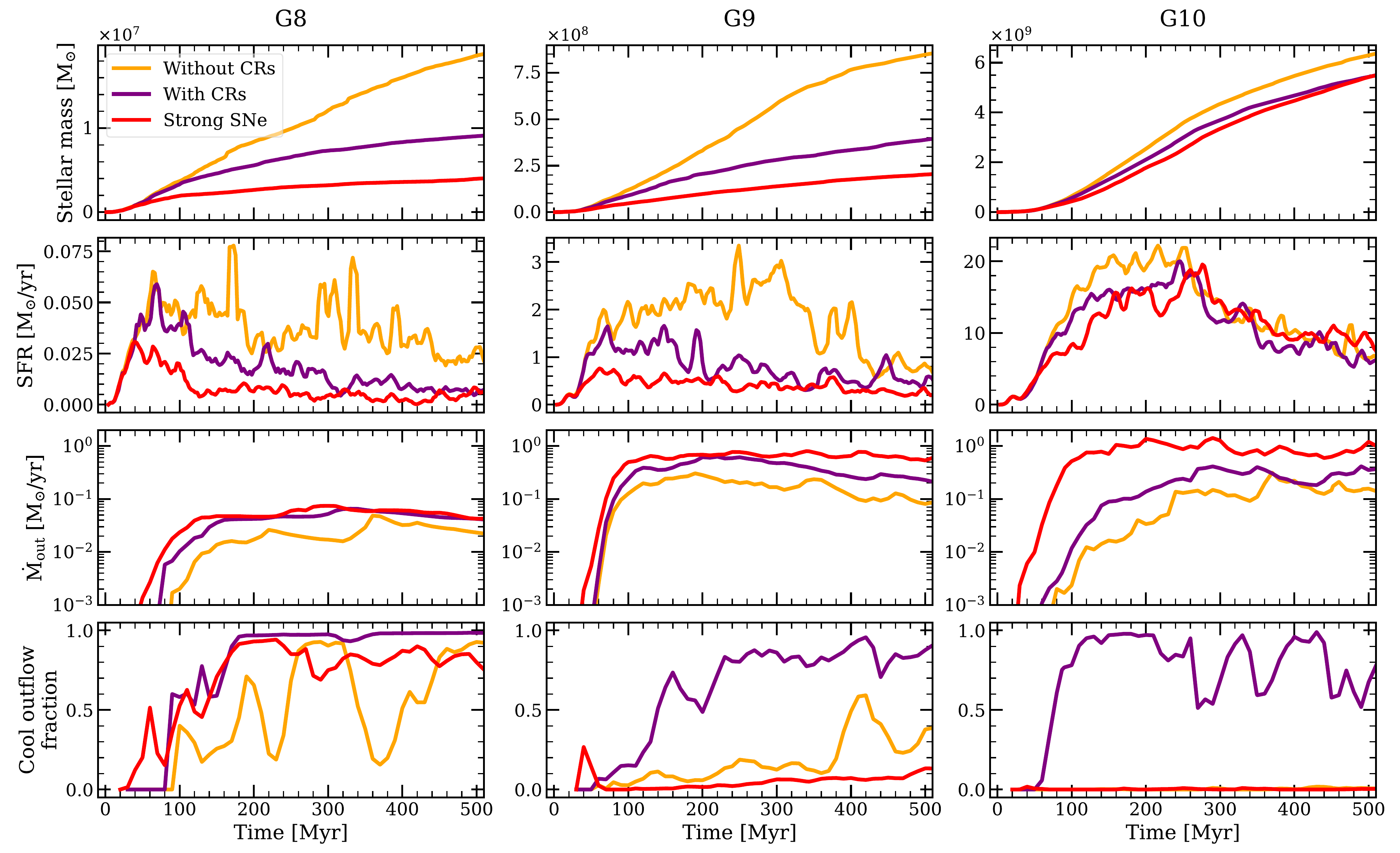}
    \caption{From top to bottom: Evolution with time of the stellar mass, star formation rate, mass outflow rate measured at 10 kpc, and fraction of outflowing gas with temperature below $10^5$ K. The orange, purple and red curves correspond to runs without CRs, with CRs ($\kappa=10^{28}\,\rm cm^{2}\,s^{-1}$) and with the strong SN feedback respectively. The strong SN feedback is somewhat more efficient than CRs in regulating star formation. It also drives stronger winds, albeit hotter than with CRs.}
    \label{fig:boost}
\end{figure*}


The objects we are focusing on in this study have a fairly low mass and a high fraction of gas (see Table~\ref{tab:run_prop}), which is typical for high-redshift galaxies \citep{Daddi2010, Tacconi2010, Tacconi2013, Genzel2015}. SNe are usually assumed to be the most efficient feedback process to regulate star formation in low-mass galaxies \citep[see e.g.][]{Dekel&Silk1986,Hopkins2011, Gelli2020}. However, they appear insufficient to explain a number of observed properties. Among others, we know from \citet{Hu2017}, \citet{Emerick2018} or \citet{Fujimoto2019} that coupling SN and radiation feedback reduces tensions between galaxy simulations and observations. All of these studies also point to the weakness of these combined feedback mechanisms to efficiently regulate star formation. To compensate the lack of efficiency of SN (and radiation) feedback in driving sufficient regulation of star formation in high-resolution simulations of galaxy evolution, various forms of sub-grid models are used, and sometimes the energy injection from supernova explosions is simply artificially boosted to reach the desired agreement with observations. This is what is done in the \sphinx{} suite of cosmological simulations \citep{Rosdahl2018}, where the number of SN explosions per Solar mass formed is amplified by a factor four in order to roughly reproduce the stellar-to-halo mass relation and the UV luminosity function at $z=6$\footnote{Nonetheless, at lower redshift ($\rm z \sim 3$), \citet{Mitchell2018} show that such an over-injection of SN energy fails and that galaxies still have too high stellar masses, indicating that complementary physics are lacking, and that even this four-fold amplification of SN feedback is not enough.}. Using here the same star formation, SN and radiation feedback implementations as in the \sphinx{} simulations, we want to assess whether CRs could be a real and physical substitute for the amplified SN feedback. In other words, we want to assess whether they provide a similar regulation of star formation when combined with un-amplified SN feedback and, if so, if their impact on the star formation operates differently, e.g. with higher or lower outflow rates or producing very different morphologies in the ISM or CGM. Thus, we perform additional isolated disc runs, labelled 'Strong SNe', where we increase the number of SN explosions per unit Solar mass by a factor four, which corresponds to SNe releasing an energy of $28.8\times 10^{48}\,\rm erg\,M_\odot^{-1}$ instead of the $7.2\times 10^{48}\,\rm erg\,M_\odot^{-1}$ derived from a canonical Kroupa IMF. We then investigate how the CR feedback (with $\rm \kappa=10^{28}\,\rm  cm^2\,s^{-1}$) compares to the calibrated boosted SN feedback adopted by \citet{Rosdahl2018} in terms of star formation regulation efficiency, outflows, and escape of LyC photons.

Figure~\ref{fig:boost} shows the differences in star formation and outflows between our galaxies with and without CRs in purple and orange, respectively, and with the strong SN feedback (and no CRs) in red. From top to bottom, we plot as a function of time the stellar mass, the SFR, the mass outflow rate measured at 10 kpc, and the fraction of mass outflow with temperature below $10^5$ K (which we term here 'cool' outflows).

Generally the strong SN feedback is more efficient than CRs, both in regulating star formation and in launching winds. Compared to strong SN feedback, CRs have a similar efficiency in suppressing star formation in our most massive galaxy, but lead to twice higher stellar masses in G8 and G9. Therefore, the inclusion of cosmic rays appears not quite sufficient to replace the effects of amplified SN feedback. Nonetheless, they provide a reasonable match for the star formation and even outflows for the lower-mass galaxies.

CRs and strong SN feedback produce very different outflow temperatures. Without CRs, the fraction of outflowing gas colder than $10^5$ K decreases drastically with galaxy mass, and the outflows are composed of hot gas only in the G10 case. While amplifying SN feedback can regulate star formation more efficiently than CRs, it affects other galaxy properties such as the CGM gas, which is almost exclusively fed by winds hotter than $10^5$ K for our two most massive galaxies (G9 and G10). Conversely, galaxies with CRs all have a significant fraction of cool outflows. These contribute to enrich the CGM with metals that trace temperatures below $10^5$ K, lacking in simulations with a strong SN feedback. For instance, by comparing HI, SiIII, SiIV and CIII CGM abundances from the COS-haloes survey \citep{Werk2013, Werk2016}, \citet{Salem2016} and \citet{Butsky2021} found that CR-driven winds can better reproduce the observed metal-enriched outflows. Based on these results, the outflowing CGM potentially provides a strong constraint on CR feedback, which we intend to investigate in future work. 
\subsection{Lyman continuum escape fraction}
\label{subsection:fesc}

Several recent simulation works find that feedback regulates the escape of ionizing radiation from galaxies \citep{Ma2016,Kimm2017, Trebitsch2017, Rosdahl2018}. Therefore, it is of a particular interest to capture the physical processes that shape galaxy evolution to understand their consequences at high-redshift, where they can play an important role in the reionization of the Universe. Cosmic ray feedback tends to smooth out density fluctuations in the ISM and generate fairly dense and cold galactic outflows. How this affects the propagation and the escape of radiation through and out of galaxies however remains unexplored. The \sphinx{} cosmological simulations \citep{Rosdahl2018} produce a reionization history which is in reasonable agreement with observations, implying that this strong feedback model produces an approximately correct mean escape fraction ($f_{\rm esc}$) of LyC radiation from galaxies. Therefore, in this section, we assess how replacing this strong SN feedback by CRs affects the escape fraction of LyC radiation from galaxies and, potentially, reionization.


\begin{figure}
    \centering
	\includegraphics[width=7.cm]{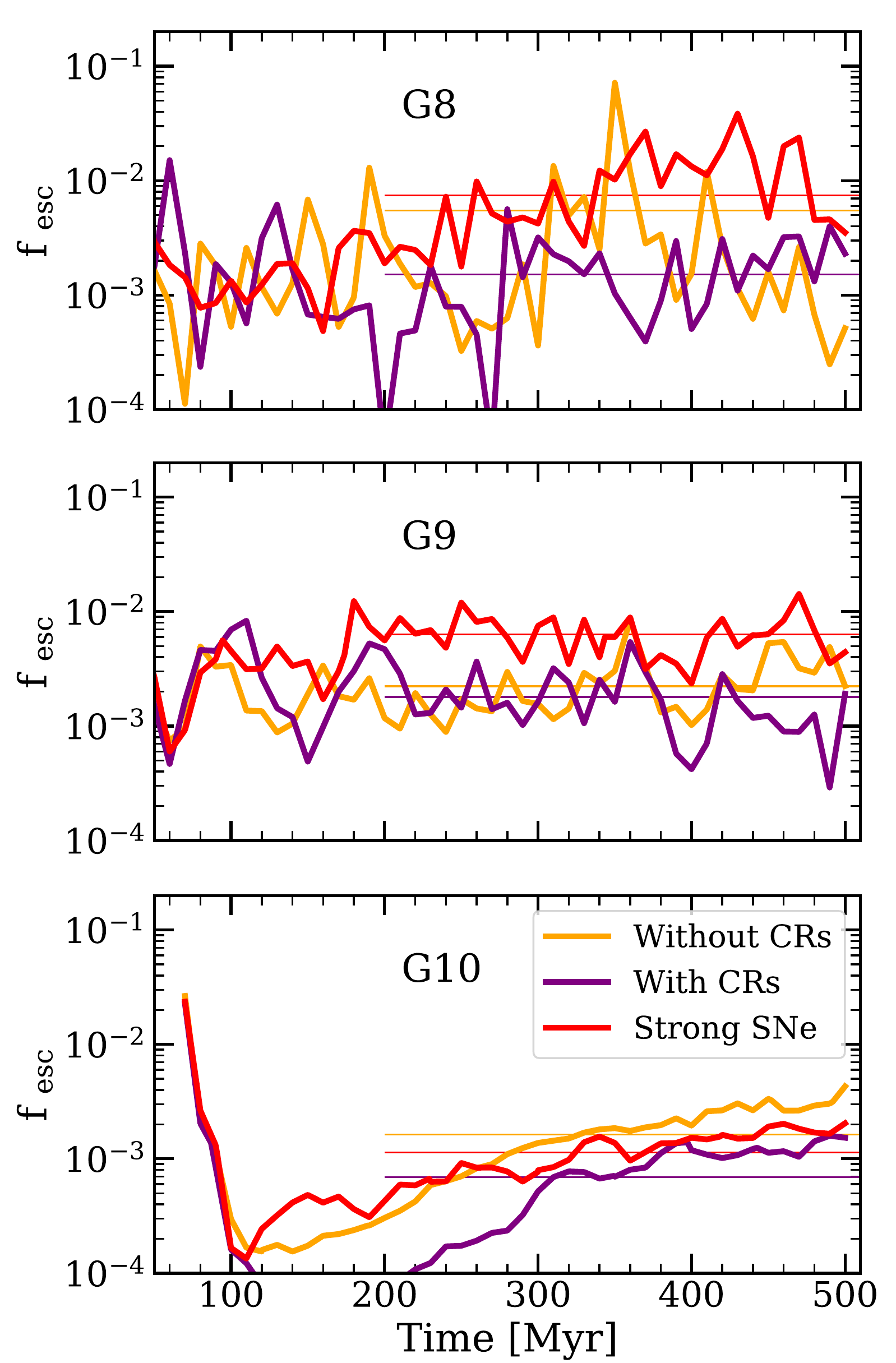}
    \caption{Escape fractions of LyC photons as a function of time. From top to bottom, panels show results for G8, G9 and G10, with (without) CRs ($\kappa=10^{28}\,\rm cm^{2}\,s^{-1}$) in purple (orange), and with the strong SN feedback in red. The thin lines correspond to the luminosity-weighted escape fractions averaged over the last 300 Myr. CR feedback consistently brings down the escape fraction by a factor of a few in our galaxies compared to the strong feedback model in \sphinx{}.}
    \label{fig:fesc}
\end{figure}
\begin{figure}
    \centering
	\includegraphics[width=7.cm]{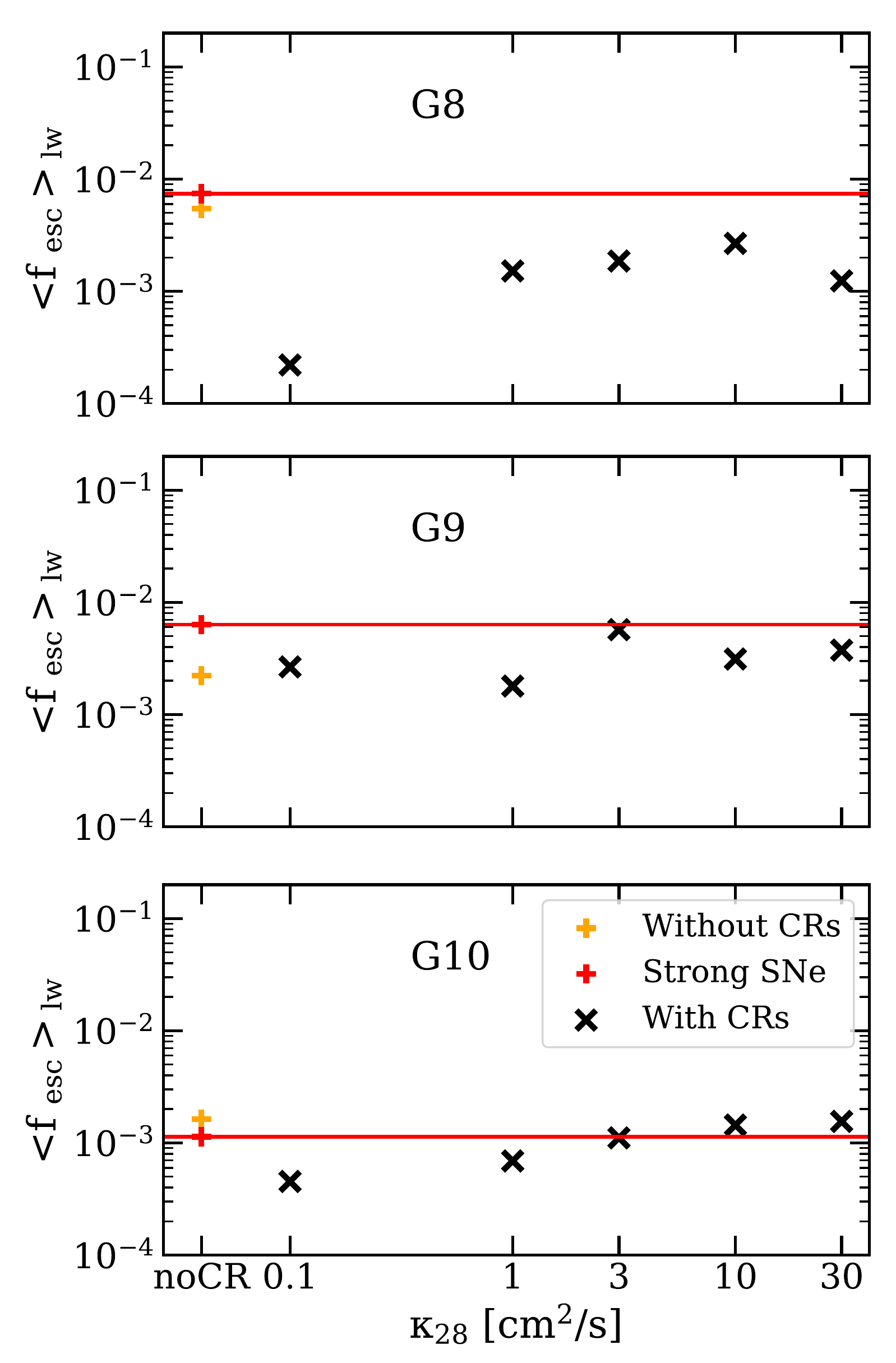}
    \caption{Luminosity-weighted escape fractions averaged over the last 300 Myr as a function of the diffusion coefficient, for increasing galaxy mass from top to bottom. The two leftmost data points show the escape fractions without CRs but with the standard and strong SN feedback in orange and red, respectively. We emphasise the escape fractions with the strong SN feedback with horizontal red lines, which we consider as the reference. Smaller diffusion coefficients suppress escape fractions, with this effect decreasing as $\kappa$ increases.}
    \label{fig:fesc-dcr}
\end{figure}
For this purpose, we estimate the LyC escape fraction in our three galaxies as follows. The escape fraction $f_{\rm esc}$ is the ratio of the photon flux measured, divided by the intrinsic luminosity emitted by the stars. For each galaxy, we estimate the total flux of all radiation groups that crosses a spherical shell of 500 pc in width, located at the viral radius $R = 41$, 89 and $192\,\rm kpc$ for G8, G9 and G10 respectively. To avoid any spurious estimation of the photon flux due to the irregular structure of the grid, we randomly sample photon fluxes from a million points inside the shell (using the \texttt{Pymses}\footnote{\href{https://irfu.cea.fr/Projets/PYMSES/intro.html}{https://irfu.cea.fr/Projets/PYMSES}} code), from which we derive an average photon flux in the shell $F_{\rm meas}$. Without any absorption, the flux would be the total intrinsic luminosity emitted by the stars $L$ divided by the area of the shell where the flux is measured. Because we use a reduced speed of light $c_{\rm red}=c/100$, it takes some time for the light to reach the shell. To correct for this delay, we compare the photon flux at a time $t$ and at a distance $R$ to the luminosity emitted a light-crossing time ago, i.e at $t-R/c_{\rm red}$. Therefore, the photon flux emitted by the stars is $F_{\rm *} = L(t-R/c_{\rm red})/(4\pi R^2)$.

The escape fraction then provides an estimate of how much radiation has escaped from the galaxy to a given distance R from its centre, such as:
\begin{equation}
    f_{{\rm esc}}(R,t) = \frac{F_{\rm meas}}{F_{\rm *}} = \frac{<c_{\rm red}N(R,t)>}{L\left(t-\frac{R}{c_{\rm red}}\right)} 4\pi R^2\, ,
	\label{eq:fesc}
\end{equation}
where $N(R,t)$ is the photon number density measured at a distance $R$ and at a time $t$.

Figure~\ref{fig:fesc} shows the escape fraction of LyC photons as a function of time for our three discs, without and with CRs in orange and purple, and with the strong SN feedback in red. The escape fraction fluctuates considerably, which is a consequence of the bursty nature of star formation and feedback, as shown by e.g., \citet{Ma2016} and \citet{Trebitsch2017}.

To provide a clearer picture, we show as thin horizontal lines luminosity-weighted average escape fraction over the last 300 Myr, in each simulation. In order of increasing disc mass, we find that the luminosity-weighted escape fractions are reduced by a factor 4.9, 3.5 and 1.6 when CRs are included compared to the case with strong SN feedback.

In the work of \citet{Rosdahl2018}, the escape fraction of LyC photons in a galaxy with a mass similar to G8 using the strong SN feedback model peaks several times at values of $\sim 0.2$ during the 1 Gyr runtime, and even reaches a maximum of 0.8. Globally, and even when using the same enhanced SN feedback, the escape fractions are much smaller in our disc galaxies. This is a limitation of idealized runs that do not reproduce the bursty and irregular galaxy growth expected at high redshift.

Regardless of their low values, we note that the escape fractions are significantly reduced with CRs. This reveals that the effects of CR feedback on thickening the galaxy discs and smoothing the ISM may have important consequences on the escape of ionizing radiation, which could spell a problem in reionization models, which already tend to struggle to produce high enough escape fractions to reionize the Universe \citep[e.g.][]{Ma2015,Ma2016}.

As the effects from CR feedback differ with the diffusion coefficient, Figure~\ref{fig:fesc-dcr} shows the luminosity-weighted escape fraction averaged over the last 300 Myr (similar as what is plotted in thin lines in Figure~\ref{fig:fesc}) with increasing diffusion coefficient from left to right, and without CRs but with a standard (in orange) and strong (in red) SN feedback for the two leftmost data points.
Because the strong SN feedback is found to produce high enough escape fractions to reionize the Universe before $z=6$ in the \sphinx{} cosmological simulations \citep{Rosdahl2018}, we emphasise with red lines the escape fraction with the strong feedback in our disc galaxies.
The escape fractions values measured in G8 with CRs are well below the strong SN feedback case, for any diffusion coefficient. However from our two higher mass galaxies, taking $\rm \kappa_{28}=3$ or higher produces similar escape fractions as the strong SN feedback. We stress that these results need to be confirmed with more realistic high-z galaxies in cosmological simulations, which tend to have much higher escape fractions than these idealised and rather structured galaxies. This will be the topic of our upcoming work.

\section{Discussion}
\label{section:discussion}

We now compare our results with those of other studies. We first focus on the effect of CRs in regulating star formation and altering the ISM, and then review their efficiency in driving winds and the dependency of this efficiency with the CR diffusion coefficient.

\subsection{CR feedback at ISM scales}
From star forming clouds to the CGM, CRs significantly affect the gas component. At ISM scales, the pressure they exert pushes the gas, which tends to smooth the overall galaxy inner gas distribution (as shown in Fig.~\ref{fig:nHmaps}). We find that the efficiency of CR feedback in directly regulating star formation weakens with increasing galaxy mass. For dwarf galaxies, we find 50\% lower SFR compared to runs without CR feedback (Fig.~\ref{fig:sfr}).

A similar reduction in star formation was found by \DDtw{}. This suggests that the effect of CRs on SF is not sensitive to the additional or different physics we include, namely more physically motivated models for star formation and SN explosions as well as the addition of radiation feedback.

The efficiency of CRs in disrupting high density regions has already been reported in other studies. The higher CR efficiency in regulating star formation in low mass galaxies has been found by e.g. \citet{Jubelgas2008}, \citet{Booth2013}, \citet{Pfrommer2016} and \citet{Wiener2017}. However, there is divergence concerning the CR feedback efficiency in galaxies as massive as G10. While our results are consistent with those of \citet{Pfrommer2016} and \citet{Buck2020}, who report very little effect of CRs on the SFR in a Milky Way mass object, they differ from those by the FIRE-2 cosmological zoom-in simulations~\citep{Chan2019}.

\citet{Chan2019} found a star formation suppression up to a factor of 1.5 from CR feedback with $\rm \kappa=3\times10^{28}\ cm^2\, s^{-1}$ in their most massive star-forming discs, whose masses are in-between those of our G9 and G10. It is unclear why CRs can still impact star formation in massive galaxies in their case and not in ours. Among the differences between our runs (in addition to the fact that they perform cosmological simulations while we study idealized galaxies), we have a local star formation efficiency with values dependent on the local gas properties, while this efficiency is set to 100\% in the FIRE simulations. This is likely to have consequences on galaxy evolution, as it directly affects the spatial and temporal distribution of star formation. Nonetheless, in agreement with the FIRE simulations, we find the same reduced CR feedback efficiency in regulating star formation with increasingly high diffusion coefficient. 

Another point to be noted in the FIRE-2 simulations is that they use a significantly higher fiducial $\rm \kappa = 3\times 10^{29}\,\rm cm^2\,s^{-1}$ to avoid CR energy losses and attain consistency with gamma-ray observations. This value yields much stronger winds and up to an order of magnitude higher mass loading factor than when CRs are excluded. In a cosmological context, as it is the case for the zoom-in simulations of \citet{Hopkins2021}, galaxies are evolved for long enough that the greater amount of outflows ends up altering significantly the gas content of a galaxy and, hence, its star formation. In our G10 galaxy, mass loading factors are not sufficiently increased to have an impact on star formation. Considering the low 0.01 to 0.1 mass loading factors in G10, much more gas is converted into stars than pushed away in the form of outflows. We can roughly estimate that a 10 times larger timescale would be needed for the outflows expelled from G10 to start impacting the SFR. By that time, most of the gas would be converted into stars. Most long-term effects due to CR driven winds extracted from our isolated simulations are somewhat speculative and require further revision using cosmological simulations, which will be the subject of future work.

\subsection{The efficiency of CRs in driving winds}
While the effects of CR feedback on star formation vary with galaxy mass, it consistently helps driving more and colder outflows, which affects the gas morphology in both the galaxy and its CGM. The non-thermal CR pressure support increases the mass loading factor by 1 dex, for all our explored galaxy masses close to the disc and at least a factor two 15 kpc away from it (Fig.~\ref{fig:eta}).

The efficiency of CRs in launching winds has been measured in a number of previous studies, e.g. \citet{Girichidis2018} in a stratified ISM, \citet{Pakmor2016} in an idealized disc, and \citet{Hopkins2021} in MW-luminosity zoom galaxies from cosmological simulations. There is broad agreement that the inclusion of CR feedback leads to colder and denser winds.

\begin{figure}
	\includegraphics[width=7.2cm]{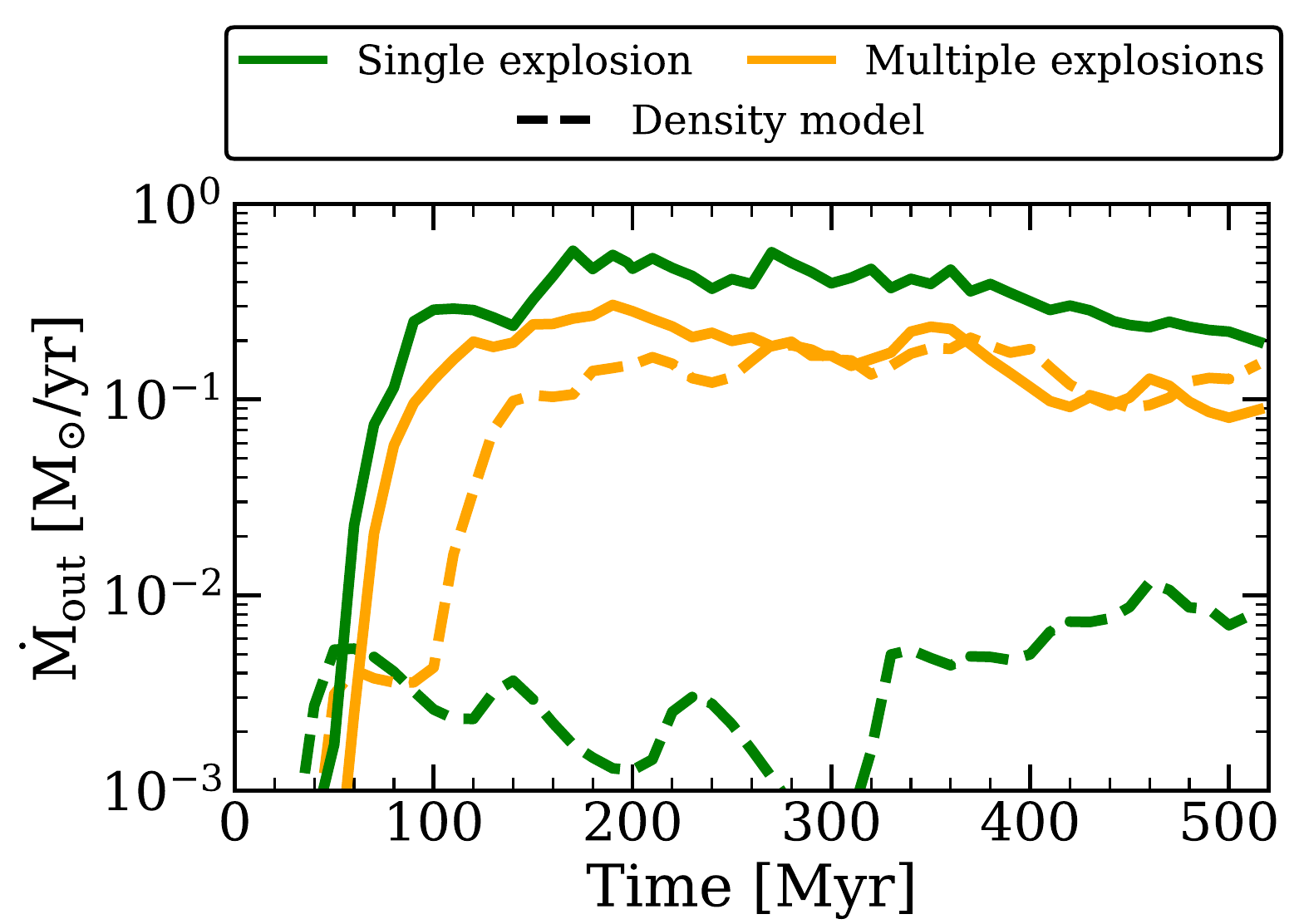}
	\centering
    \caption{Mass outflow rate versus time measured at 10 kpc from the G9 galaxy. Orange (green) curves show runs with multiple (single) SN explosions. Solid lines correspond to runs with our turbulent star formation model, while dashed correspond to the model based on the gas density used by \DDtw{}. Therefore, the dashed green curve is our closest equivalent to the setup adopted by \DDtw{}. Only switching to both the turbulent SF model and multiple explosions can explain the higher outflow rate we measure for the same galaxy.}
    \label{fig:outf_goharlike}
\end{figure}

However, it is difficult to quantitatively compare the effects of CRs from one simulation to another because of the different feedback models used. We noted in Section~\ref{subsection:cgm} that we measure a smaller enhancement of the outflow rate when adding CRs compared to \DDtw{}. This can be explained by the fact that they measure one order of magnitude lower outflow rates in their G9 without CRs than we do. This last aspect is not due to the inclusion of radiative transfer, as \citet{Rosdahl2015} showed that radiation pressure and photo-heating have a negligible impact on wind launching (and we have confirmed this in our simulations). To determine the reason for the higher mass loading factor in our G9 compared to \DDtw{}, Figure~\ref{fig:outf_goharlike} shows the outflowing gas versus time, measured at 10 kpc from G9\_noCR, comparing our feedback and star formation models against those of \DDtw{}. We show the star formation model using a density threshold and a small constant star formation efficiency used by \DDtw{} in dashed lines (that we label 'density model') and we show the turbulent model we adopt in solid lines. We also show in orange the multiple SN explosions per particle model we use compared to the single per particle model used by \DDtw{} in green.

With the density SF model (dashed), switching from one (in green) to multiple explosions per stellar particle (in orange) increases the outflow rate by up to one order of magnitude. This difference corresponds to that noted between the outflows measured by \DDtw{} and measured in this paper. At first glance, it may appear that the SN model is the factor governing the outflow rate. However, if we focus on the turbulent SF model (in solid lines), switching from one to multiple SN explosions leads to the opposite trend, with slightly stronger outflows in the former case. With the density SF model, stars are more broadly distributed as the gas to stars conversion can only occur if the gas density exceeds a certain threshold. If a stellar particle explodes only once, the gas is disrupted locally, but this single event is not enough to launch significant winds. A subsequent disruptive event is needed to take advantage of the previous one and make it easier to drive gas out of the disc. This is exactly what happens when switching from single to multiple SN explosions with the density star formation model. Conversely, the turbulent model leads to more bursty and clumpy star formation. Because stars form in more localised clumps, they also explode in very rapid succession, even if each particle explodes only once, so multiple explosions do not offer the same advantage as in the density model where star formation is more scattered in time. 

All in all, the effects of varying the locality and burstiness of star formation and subsequent SN explosions is somewhat unpredictable and non-linear, as has previously been reported in e.g. \citet{Keller2020,Andersson2020, Smith2021}. Consequently, the large difference in outflow rate measured by \DDtw{} and in this paper can only be explained by the non-linear interplay of the combination of a more bursty and clumpy SF model with multiple individual $10^{51}$ erg SN explosions for each stellar particle. With our setup, the star formation sites tend to be destroyed by the first SNe, letting later SN explosions take place in a more diffuse medium, where more momentum can be generated. As a consequence, our SN feedback is more efficient in driving outflows, and the added effect of CRs becomes smaller. Besides, we show in Appendix~\ref{app:dens-fk2} that changing from the density to the turbulent SF model with the same setup otherwise leads to different outflow rates exclusively for G8, indicating that there are additional factors that play a role, such as particle and cell resolution.

\subsection{Outflow rates and the diffusion coefficient}
The ejection of gas can significantly differ depending on the CR transport mechanism. For this reason, we investigate the role of the diffusion coefficient, one of the key parameters controlling CR feedback efficiency. As \DDtw{} with G8 and G9, we find that the higher the diffusion coefficient, the higher the outflow rate (with the exception of our largest diffusion coefficient value for which the outflow rate stagnates or even slightly decreases). This is due to a more efficient escape of CRs from the disc, and hence more energy in the CGM to drive outflows. In agreement with most works\citep[see e.g.][]{Salem&Bryan2014,Jubelgas2008,Farber2018,Chan2019}, we find that a faster diffusion leads to more star formation. However, \cite{Salem&Bryan2014}, \cite{Jacob2018}, \cite{Girichidis2018}, and \cite{Quataert2021} all find higher mass loading factors with a lower diffusion coefficient, while we measure the opposite trend. We discuss below the reason behind this discrepancy.

Figure~\ref{fig:timescales} suggests that CRs are more efficient to drive winds if they quickly escape the ISM where CR energy losses are dominant, or equivalently if more CR energy remains to push winds away from the galaxies with a high diffusion coefficient. In simulations without CR energy losses, a low diffusion coefficient allows CRs to escape from the disc slowly enough to drive more outflows, without losing the energy needed to accomplish this during the time they are confined in the galaxy. This is the case for the $\rm 10^{12}\, M_\odot$ halo from \citet{Salem&Bryan2014}, where CR radiative losses are not included. In our case, we clearly see from Figure~\ref{fig:DCRmaps} that the CR pressure gradient cannot build up to large distances around G10 when $\rm \kappa = 10^{27}\,\rm cm^2\,s^{-1}$. For this value, CRs remain trapped close to the disc, where they lose all their energy. Furthermore, as the aforementioned studies find, we start to see a hint that with the extremely high $\rm \kappa = 3\times10^{29}\,\rm cm^2\,s^{-1}$, increasing the diffusion coefficient towards very high values make CRs escape so quickly that their effect starts to vanish.

While reaching a similar resolution as we have and including CR energy losses, \citet{Girichidis2018} found slightly stronger outflows with decreasing coefficient values in their stratified box of ISM. However, they measure outflows at 1 and 2 kpc. At these closer distances to the mid-plane, it is difficult to distinguish between CGM outflows and ISM fountains. In Appendix~\ref{app:out-2kpc}, we show that we do measure higher outflow rates with higher $\kappa$ at 2 kpc from the galaxies, but the differences with varying diffusion coefficient become much smaller than at 10 kpc.

Interestingly, \citet{Jacob2018} found in agreement with our results that the diffusion coefficient for which the maximum of outflows is reached varies with galaxy mass. This is consistent with a critical diffusion coefficient value below which the wind properties change, such as its velocity as found by \citet{Quataert2021} and its temperature as we show in Figure~\ref{fig:outflows-DCR}. All of this implies that the exact behaviour of outflows with changing diffusion coefficient depends on multiple parameters, from the initial conditions of the galaxy (e.g. its size) to the star formation and feedback sub-grid models.

Finally, we remark that CR propagation does not really occur under a constant diffusion coefficient. This motivates our study of the effects of CR feedback under different $\kappa$ values. Ideally, one would couple the dependency of the diffusion coefficient with CR particle energy and with the gas ionization state (as CRs are charged particles that are more tightly coupled with a fully ionized gas), as well as accounting for CR streaming. The energy and ionization state dependency of the diffusion coefficient have been independently studied by \citet{Girichidis2021} and \citet{Farber2018} respectively, who both showed a greater CR feedback efficiency in regulating star formation and driving winds. Improved CR propagation models are therefore crucial to improve our understanding of the role of CRs on galaxy evolution.

\subsection{Dependency of the results with resolution}

\begin{figure}
	\includegraphics[width=\columnwidth]{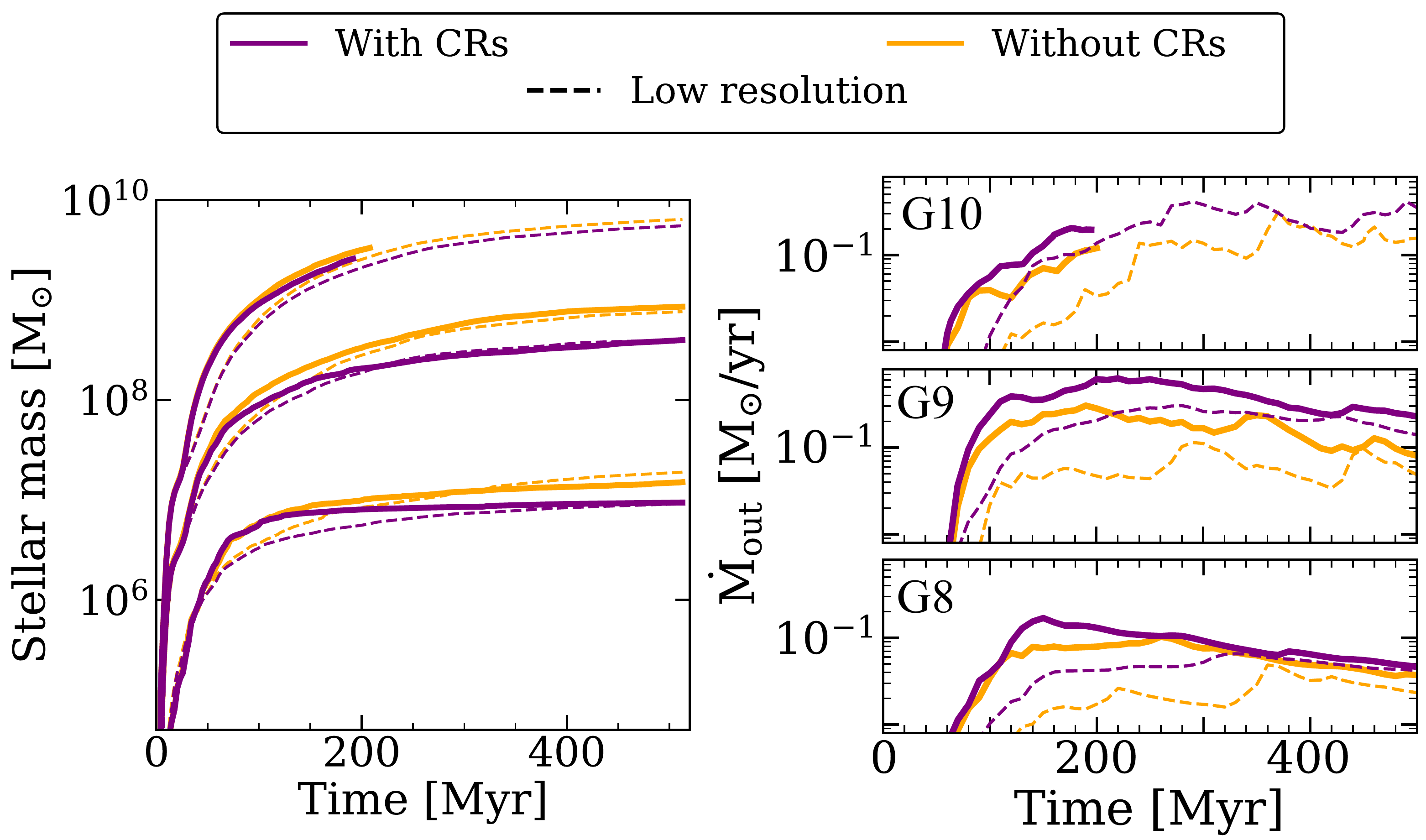}
	\centering
    \caption{Stellar mass (left panel) and mass outflow rate at 10 kpc from the disc (right panels) as a function of time for G8, G9 and G10 from the lower to the upper panels. The orange curves correspond to the galaxies without cosmic ray feedback, and the purple to runs with CRs added. Solid lines represent the runs with a fine resolution of 9 pc for G9 and G10 and 4.5 pc for G8, while the dashed lines show the equivalent with a 18 pc maximum resolution for the two more massive galaxies and 9 pc for G8.}
    \label{fig:res}
\end{figure}

Finally, we briefly discuss how our results vary with resolution.
In Fig~\ref{fig:res}, we show the evolution with time of the stellar mass formed (left panel) and the outflow rate at 10 kpc from the midplane of the disc (right planel). We compare low resolution runs, where the minimum cell width is 9 pc for G8 and 18 pc for G9 and G10 (dashed thin lines) to runs where the minimum cell width is 4.5 pc for G8 and 9 pc for the two other galaxies (solid lines). In the low resolution runs, the maximum cell width is 2.34 kpc, 4.68 kpc and 4.68 kpc and the stellar particle mass is $2.5\times 10^3\ \rm M_\odot$, $2\times 10^4\ \rm M_\odot$ and $2\times 10^4\ \rm M_\odot$ by increasing order of galaxy mass. In the high resolution runs, the maximum cell width is 2.34 kpc for all the galaxies and the stellar particle mass is $310\ \rm M_\odot$, $2.5\times 10^3\ \rm M_\odot$, and $2.5\times 10^3\ \rm M_\odot$ by increasing order of galaxy mass.

During the first 70 Myr, the runs with the higher resolution tend to produce more stars, regardless of whether CRs are included or not. However, after the initial collapse of the disk and once star formation stabilises (by $t \sim 200$ Myr), the stellar masses are similar, with approximately equal final stellar masses regardless of resolution. Moreover, we find that the CR feedback efficiency in suppressing star formation and driving winds does not depend sensitively on the cell resolution.
Regarding the right panel, we systematically measure stronger outflows at higher resolution. Gas reaches higher densities in runs with higher resolution, and this may lead to more efficient entrainment of gas into galactic winds. Yet we approximately measure the same increase of outflows when CR feedback is included, with the noticeable exception of G8. We further note that the mass outflow rate in G8 becomes similar in the last 150 Myr, for the low and high resolution runs. We also emphasise the fact that the temperature of these outflows follows the same trend regardless of the resolution (not shown in the paper), with runs with CRs being always dominated by warm and cold gas while runs without CR are almost entirely hotter than $10^5$ K. Therefore, even though varying the resolution changes quantitatively some of our results, it does not impact the strength of CR feedback compared to the SN feedback alone (boosted or not). Changing the resolution does not either impact the effects of CRs in producing a colder CGM and impacting the escape of ionising radiation, and is unlikely to change the effects of varying the diffusion coefficient.


\section{Conclusions}
\label{section:ccl}

The aim of this study is to investigate the role of CRs in suppressing star formation in galaxies and driving outflows. For this purpose, we perform the first cosmic ray radiation-magnetohydrodynamics simulations of three gas-rich galaxies of different masses, using the \ramsesrt{} code \citep{Teyssier2002, Rosdahl2013}, merged with the magnetohydrodynamics implementation of \citet{Fromang2006} and modified to include anisotropic cosmic ray transport as described by \citet{Dubois&Commercon2016}. By comparing our three galaxies with and without CRs, added to our fiducial SN and ionizing radiation feedback, we first investigate how the effects of CR feedback vary with galaxy mass. However, the uncertainty associated with CR transport (specifically the value of the CR diffusion coefficient) complicates determining these effects. For this reason, we investigate in more detail how our results change when varying the diffusion coefficient from $10^{27}$ to $\rm 3\times10^{29}\,\rm cm^2\,s^{-1}$, which are reasonable limits for the CR diffusion. We also study to what extent CRs provide the galaxy growth suppression required by cosmological simulations. We compare the efficiency of CRs in regulating star formation against the same calibrated model with enhanced SN strength employed by the \sphinx{} simulations of reionization \citep{Rosdahl2018}. This serves as a precursor of our future work reviewing this in cosmological simulations. In addition, this allows us to assess how CR feedback affects the escape of LyC radiation from galaxies, and hence, indirectly, the process of reionization. We summarise our main conclusions as follows.

\begin{itemize}
\item \textit{Cosmic rays have an important effect on the ISM.} On ISM scales, the pressure from cosmic rays tends to smooth out density contrasts. This also tends to produce thicker gas disks than without CR feedback.
\item \textit{CR feedback efficiency in regulating star formation decreases with galaxy mass.} In our two dwarf galaxies, for which the gravitational potential is relatively weak, CRs can easily act locally. They make the regions where stars form and explode more diffuse, which reduces in turn the number and the mass of star-forming clumps. However, for galaxies so massive that even SN feedback starts being inefficient, as it is the case for G10, they have almost no effect at ISM scales.
\item \textit{At any galaxy mass, CRs drive stronger and colder outflows than thermal pressure from SNe and radiation alone.} Depending on the distance where the outflows are measured, CR feedback increases the mass loading factor by a factor of 2-10. With CRs, the outflows are much colder, predominantly between $10^4$ K and $10^5$ K, and we measure outflows colder than $10^4\,\rm K$, completely absent in runs without CRs.
\item \textit{Low diffusion coefficients make CRs act locally, smoothing out density contrasts in the ISM and reducing star formation, but having negligible and even negative effects on outflow rates.} The regulation of star formation depends on the amount of CR energy trapped in the disc. With low diffusion coefficient, CRs remain confined longer, thus having more time to interact with the ISM gas and therefore decrease the star formation rate, but cannot drive strong outflows. 
\item \textit{The mass outflow rate and temperature composition are sensitive to the diffusion coefficient.} The higher the diffusion coefficient, the greater the outflow rate, with a consistently higher fraction of hot gas. We find an inflexion value that depends on galaxy mass, beyond which the amount of cold outflows drops, and the total mass outflow rate stagnates. Although we do not capture it in our simulations, if the galaxies were to evolve for a longer time and in a cosmological context, the large increase in the mass-loading factor found for our lower-mass galaxies is likely to lead to a long-term suppression in star formation.
\item \textit{CR feedback does not provide a sufficient 'replacement' for artificially enhanced SN feedback model used in high-redshift cosmological simulations.} The strong SN model is especially more efficient than CRs to regulate star formation. In addition, the outflows driven by SN or CR feedback have very different temperatures. While the CR energy is linked to the ability of pushing more dense and cold gas from the ISM, the SN feedback, boosted or not, tends to push only hot and diffuse gas from the galaxies.
\item \textit{Replacing the strong SN feedback used in the \sphinx{} cosmological simulations of reionization by CR feedback reduces the escape fraction of LyC radiation significantly.} 
\end{itemize}

Overall, we find CR feedback to notably impact star formation in low-mass galaxies, and it alters the amount and the temperature of the outflowing gas. The quantification of these effects is however sensitive to the diffusion coefficient and, comparing to other works, to the details of initial conditions and sub-grid models. It is also important to consider the limitations of idealized non-cosmological simulations. While they provide the perfect laboratory to explore the secular effects of CR physics and their interplay with our sub-grid models, they model a highly unrealistic circum-galactic medium, which may impact the properties of the feedback-driven outflows. Besides, they are not evolved long enough to consistently capture the consequences of gas ejection, and overlook the effects of gas inflows. Cosmological simulations are required in order to consistently predict the consequences of CR-driven winds on long term galaxy evolution. In a follow-up paper, we will study CR feedback in cosmological zoom simulations using the same methods and physics employed here. We will then have a better picture of the efficiency of CRs in shaping high-redshift galaxies and the escape of ionizing photons, in order to determine how cosmic ray feedback may affect reionization.

\section*{Acknowledgements}

The authors thank the anonymous referee for the constructive comments which improved the manuscript. We gratefully thank Dmitry Makarov for his careful reading of the paper and his comments, as well as Benoît Commerçon, Léo Michel-Dansac, Edouard Tollet and Gohar Dashyan for insightful discussions. We thank Taysun Kimm for making the methods for star formation and feedback available and for helping to set up the simulations. This work has been granted access to the HPC resources of TGCC under the allocation 2020-A00806955 made by GENCI. A consequent part of the runs were performed at the Common Computing Facility (CCF) of the LABEX Lyon Institute of Origins (ANR-10-LABX-0066). We also acknowledge support from the PSMN (Pôle Scientifique de Modélisation Numérique) of the ENS de Lyon for the computing resources.\\

\textsl{Software}: \numpy{} \citep{VanderWalt2011}, \matplotlib{} \citep{Hunter2007}, \ramses{} \citep{Teyssier2002}, \pymses{}

\section*{Data Availability}
The data underlying this article will be shared on reasonable request to the corresponding author.


\DeclareRobustCommand{\VAN}[3]{#3}
\bibliographystyle{mnras}
\bibliography{biblio} 




\appendix

\section{Density threshold vs more realistic thermo-turbulent star formation models}
\label{app:dens-fk2}

\begin{figure}
	\includegraphics[width=7.2cm]{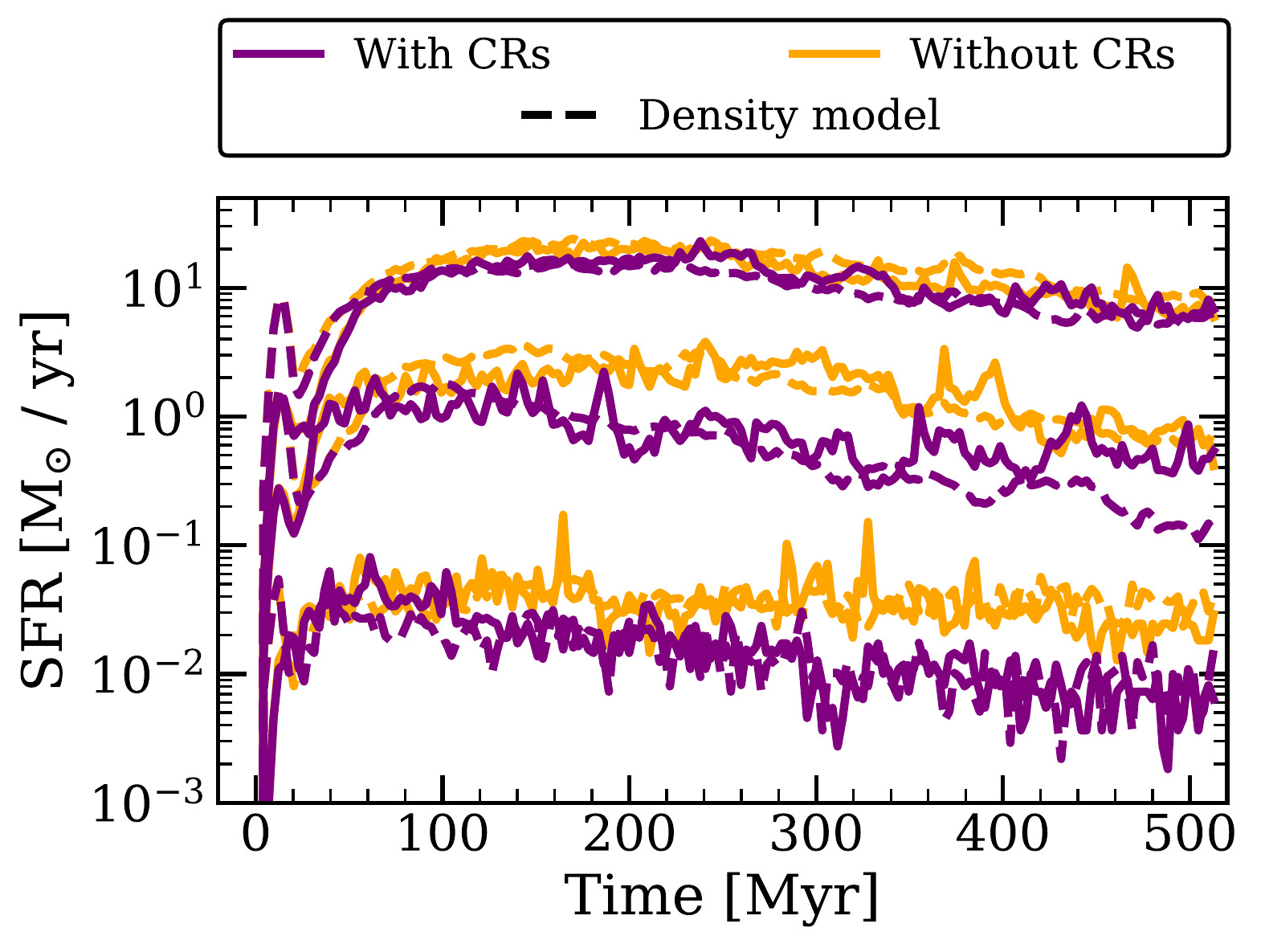}
	\centering
    \caption{Star formation rate versus time for G8, G9 and G10 from bottom to top. Orange (purple) curves show runs without (with) CRs. Solid lines correspond to runs with the turbulent star formation model, while dashed are for the classical model based on a gas density threshold and a constant but small local star formation efficiency. The two SF models produce fairly similar star formation at any galaxy mass, though it tends to be more bursty with the turbulent model. The density SF model tends to slightly increase cosmic ray feedback efficiency, especially for G9.}
    \label{fig:sfr_dens-fk2}
\end{figure}

\begin{figure}
	\includegraphics[width=7.2cm]{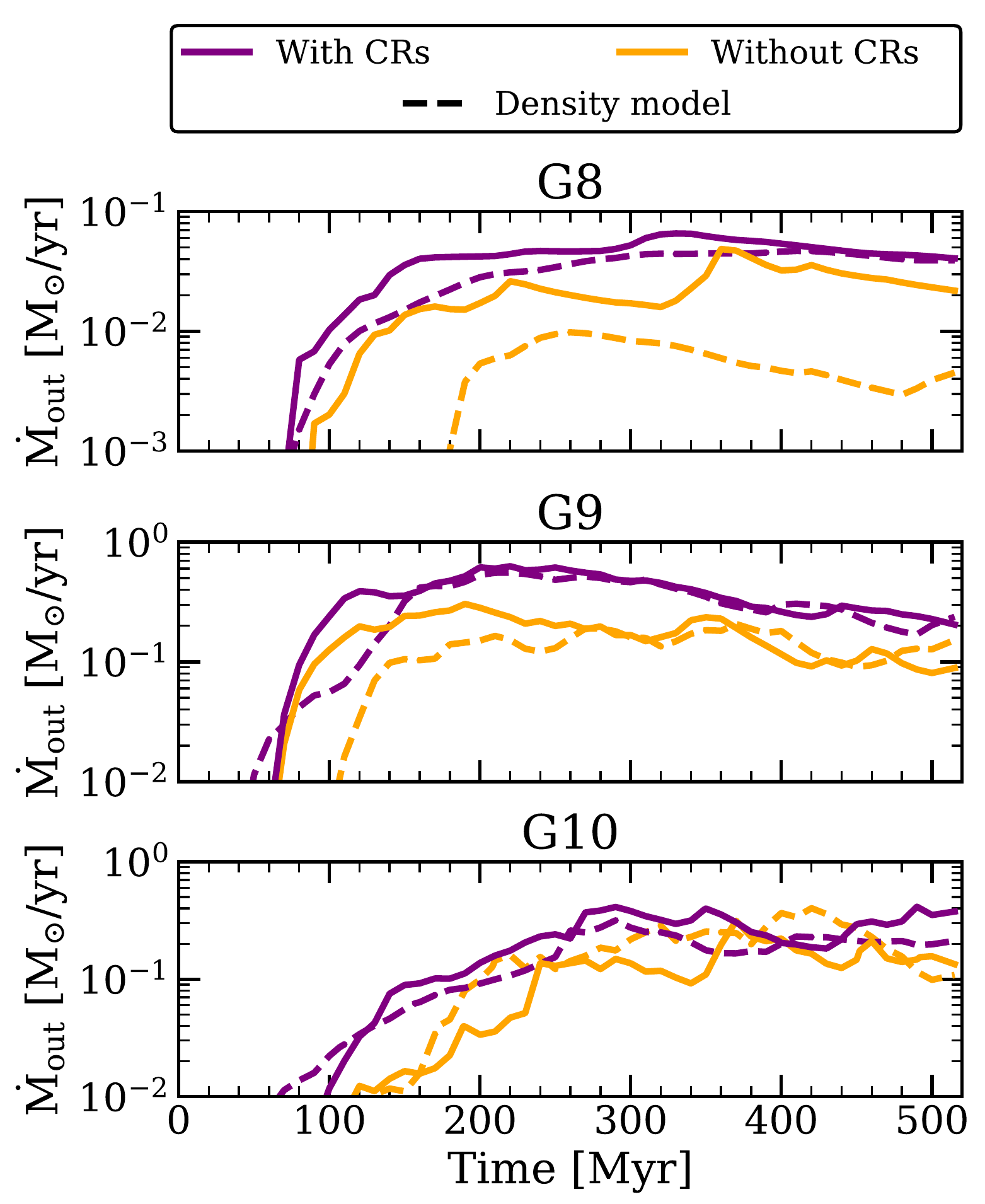}
	\centering
    \caption{Mass outflow rate versus time measured at 10 kpc from the discs, by increasing order of galaxy mass from top to bottom. Orange (purple) curves show runs without (with) CRs. Solid lines correspond to runs with the turbulent star formation model, while dashed are for the classical density threshold model. The two SF models give similar results at any galaxy mass and regardless of the inclusion of CRs, except for a significantly enhanced amount of outflowing gas for G8 without CRs when we switch from the density to the turbulent SF model, as discussed in Section~\ref{section:discussion}.}
    \label{fig:outf_dens-fk2}
\end{figure}

This paper aims to provide a continuity to the results presented by \DDtw{}. One of the main differences between their setup and the one adopted in this work is the star formation model, as described in Section~\ref{subsection:SF}. Because the SF model employed by \DDtw{} is based on a density threshold criterion, we refer to this one as "density". In our fiducial runs, the formation of stars is based on the gravo-turbulent properties of the gas, taking inspiration from \citet{Federrath&Klessen2012} work, hence the SF model is labelled "turbulent".

We investigate how sensitive our results are to the star formation model. Fig~\ref{fig:sfr_dens-fk2} show star formation rate versus time for our three galaxies, with (without) CRs in purple (orange). The solid lines show the same as in Fig.~\ref{fig:sfr}, that is to say galaxies run with the turbulent model, while the dashed lines show the counterparts with the density SF model. We find that the turbulent SF model is generally more bursty than the density model. Globally, the stellar mass formed with one model or another is almost the same, at any time and no matter the galaxy mass. The small differences, especially during the last 200 Myr for G9, suggest that CRs are more efficient to reduce the star formation with the density SF model.

Figure~\ref{fig:outf_dens-fk2} shows the mass outflow rate as a function of time, measured at 10 kpc from the three galaxies. We apply the same colour code as for Fig~\ref{fig:sfr_dens-fk2} to distinguish between the inclusion of CRs or not and the star formation model. With the exception of G8, the same flux of outflowing gas is measured for both models after a few hundred Myr. Only G8 without CRs has considerably higher outflow rates with the turbulent SF model than with the model using a gas density threshold, and consequently shows a less significant contribution of CRs, as discussed in Section~\ref{section:discussion}.

\section{Outflows at 2 kpc with different diffusion coefficients}
\label{app:out-2kpc}

\begin{figure}
    \centering
	\includegraphics[width=7.2cm]{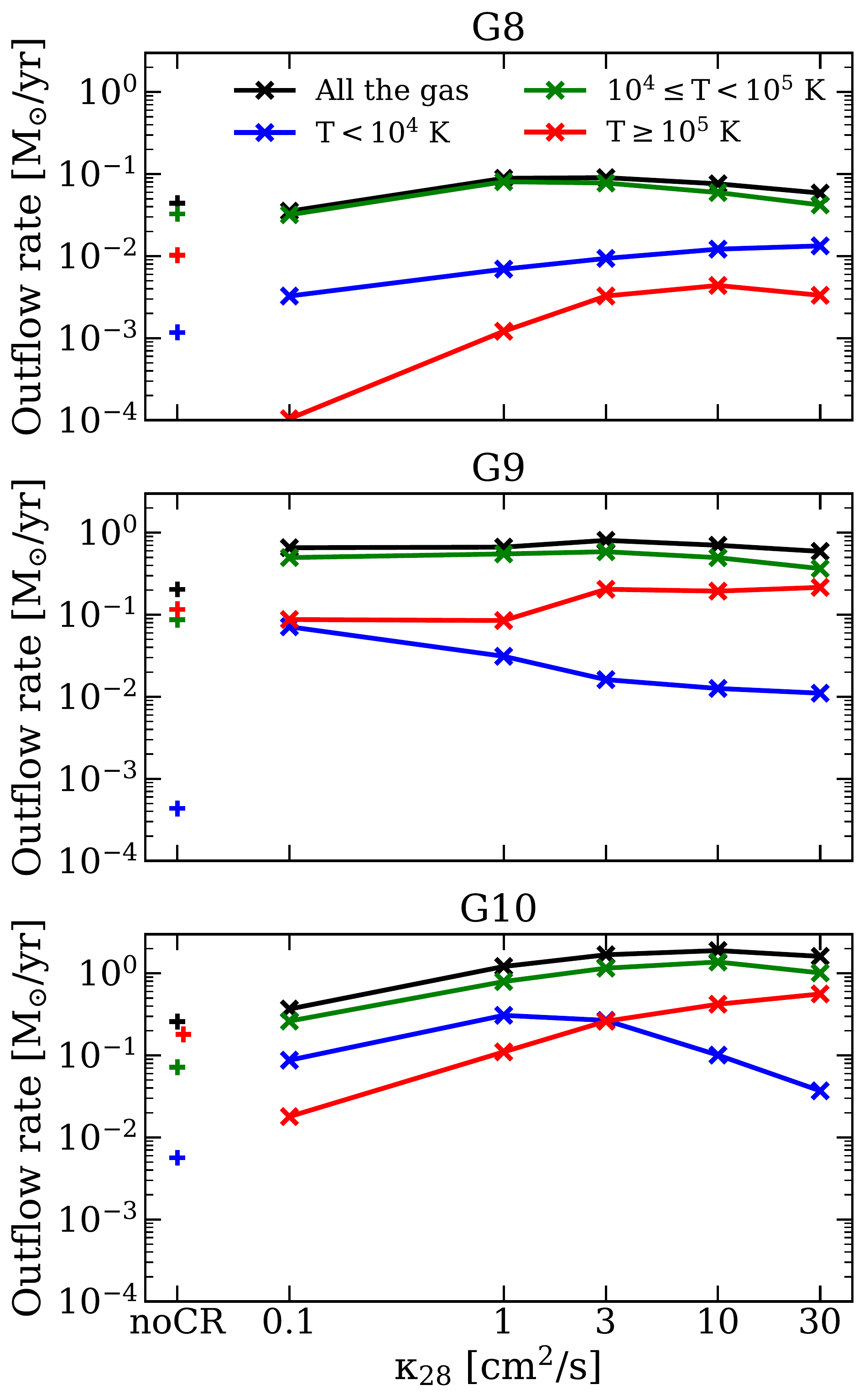}
    \caption{Gas outflow rate at 2 kpc from the galaxy midplane as a function of the diffusion coefficient, in order of increasing galaxy mass from left to right. For each galaxy, data are stacked between 200 and 500 Myr. We show the total amount of outflowing gas in black, the cold ($\rm T < 10^4\ K$) in blue, the warm ($\rm 10^4 \leq T < 10^5\ K$) in green and the hot outflows ($\rm T \geq 10^5\ K$) in red . The leftmost symbols represent galaxies without CRs. Similarly to outflows measured at 10 kpc, the total rate of outflowing gas and especially its hot component are globally enhanced with higher values of $\kappa$. However, the rate of cool outflows stops increasing and even drops, at a diffusion coefficient limit which increases with galaxy mass.}
    \label{fig:outflows-DCR2}
\end{figure}

Figure~\ref{fig:outflows-DCR2} is the equivalent of Fig.~\ref{fig:outflows-DCR} for mass outflow rate measured at 2 kpc (instead of 10 kpc for the latter). We show the outflows for increasing galaxy mass from left to right as a function of the diffusion coefficient, at the exception of the leftmost symbols that represent runs without CR feedback. For each simulation we take the average of 31 snapshots (with 10 Myr intervals) between 200 and 500 Myr to compute the outflow rates, shown in black, blue, green and red for the total, cold, warm and hot outflowing gas respectively. Generally, the trends are similar at 2 kpc and at 10 kpc. We measure increasing hot outflow rate with higher diffusion coefficient, and warm and cold outflows that stagnate or even decrease above a given value of $\kappa$ which differs with galaxy mass. While at 10 kpc, G10 has more outflows without CRs than with the lowest $\rm \kappa=10^{27}\,\rm  cm^2\,s^{-1}$, the total outflow rates are slightly higher with the small diffusion coefficient than without CRs at 2 kpc. This is also visible in Fig.~\ref{fig:DCRmaps}, where using $\rm \kappa=10^{27}\,\rm  cm^2\,s^{-1}$ produces a thick disc with dense gas close to the galaxy midplane, but remains inefficient to push the outflows further out as most of the CR energy is likely dissipated before escaping the galaxy. Besides, CRs diffusion is slower with lower diffusion coefficient, as visible in Fig.~\ref{fig:timescales}. The winds supported by CRs that have $\rm \kappa=10^{27}\,\rm  cm^2\,s^{-1}$ are too slow to be equally measured at small and large distances from the galaxy, as they quickly fall back to the disc if they cannot escape its gravity rapidly enough.

\bsp	
\label{lastpage}
\end{document}